\documentclass[12pt]{article} 
\usepackage[a4paper,textwidth=490.0pt,textheight=703.1pt]{geometry}
\usepackage{mathtools}
\usepackage{subcaption}
\usepackage{tikz}
\usetikzlibrary{decorations.pathreplacing,decorations.markings}
\usepackage{hyperref}

\usepackage{eucal}
\usepackage{url}
\usepackage{booktabs}
\usepackage{epsfig} %
\usepackage{float} %
\usepackage{amssymb} %
\usepackage{amsmath} %
\usepackage{verbatim} %
\usepackage{cite} %
\usepackage{color} %
\usepackage{eufrak}
\usepackage[english]{babel}
\usepackage[latin1]{inputenc}
\usepackage[T1]{fontenc}
\usepackage{ae}
\usepackage{url}   
\usepackage{graphicx} 
\usepackage{booktabs}
\usepackage{slashed}

\usepackage[capitalize]{cleveref}

\newcommand{\I}{\mathrm{i}}
\DeclareMathOperator*{\Disc}{Disc}

\newcommand{\F}{{\sf F}}
\newcommand{\gx}{{\tilde{g}}}
\newcommand{\GA}{{\rm G}}
\newcommand{\HA}{{\rm H}}

\newcommand*\pFqskip{8mu}
\catcode`,\active
\newcommand*\pFq{\begingroup
        \catcode`\,\active
        \def ,{\mskip\pFqskip\relax}%
        \dopFq
}
\catcode`\,12
\def\dopFq#1#2#3#4#5{%
        {}_{#1}F_{#2}\biggl[\genfrac..{0pt}{}{#3}{#4};#5\biggr]%
        \endgroup
}

\newcommand{\rmd}{\mathrm{d}}

\newcommand{\ep}{\varepsilon}

\newcommand{\Li}{\mbox{Li}}

\newcommand{\Mvec}{\mbox{\rm\bf M}}

\newcommand{\beq}{\begin{equation}}
\newcommand{\eeq}{\end{equation}}
\newcommand{\bea}{\begin{eqnarray}}
\newcommand{\eea}{\end{eqnarray}}

\allowdisplaybreaks[4]

\begin{document} 
\setlength{\baselineskip}{0.515cm}

\sloppy 
\thispagestyle{empty} 
\begin{flushleft} 
DESY 20--053 
\\ 
DO--TH 23/01
\\ 
CERN-TH-2023-020
\\ 
ZU--TH 13/23
\\
March 2023 
\end{flushleft}

\mbox{} \vspace*{\fill} \begin{center}

{\Large\bf The inverse Mellin transform via analytic continuation}

\vspace{3cm} 
\large 
{\large A.~Behring$^a$, J.~Bl\"umlein$^b$ and K.~Sch\"onwald$^c$ }

\normalsize 

\vspace{1.cm} 
{\it $^a$Theoretical Physics Department, CERN, 1211 Geneva 23, Switzerland}

\vspace{2mm} 
{\it $^b$Deutsches Elektronen--Synchrotron DESY, Platanenallee 6, 15738 Zeuthen, Germany}

\vspace*{2mm} 
{\it $^c$Physik--Institut, Universit\"at Z\"urich, Winterthurerstrasse 190, CH-8057 Z\"urich, Switzerland}


\end{center} 
\normalsize 
\vspace{\fill} 
\begin{abstract} 
\noindent
We present a method to calculate the $x$--space expressions of massless or massive operator 
matrix elements in QCD and QED containing local composite operator insertions, depending on 
the discrete Mellin index $N$, directly, without computing the Mellin--space expressions in 
explicit form analytically. Here $N$ belongs either to the even or odd positive integers. The 
method is based on the resummation of the operators into effective propagators and relies on 
an analytic continuation between two continuous variables. We apply it to iterated integrals 
as well as to the more general case of iterated non--iterative integrals, generalizing the 
former ones. The $x$--space expressions are needed to derive the small--$x$ behaviour of the 
respective quantities, which usually cannot be accessed in $N$--space. We illustrate the method 
for different (iterated) alphabets, including non--iterative $_2F_1$ and elliptic structures, 
as examples. These structures occur in different massless and massive three--loop calculations. 
Likewise the method applies even to the analytic closed form solutions of more general cases of 
differential equations which do not factorize into first--order factors. 
\end{abstract}

\vspace*{\fill} \noindent
\newpage

\section{Introduction} 
\label{sec:1}

\vspace*{1mm} 
\noindent
Precision measurements of important observables in QCD and QED \cite{Politzer:1974fr,Geyer:1977gv,
Buras:1979yt,Reya:1979zk,Blumlein:2012bf} require precision predictions through higher--order 
corrections. 
For many measurements at current and future colliders these are corrections
up to three--loop order or even higher. The application of these corrections to the data allows precision 
predictions of fundamental parameters of the Standard Model of elementary particle physics. The 
underlying theoretical
calculations require the development of efficient technologies to calculate the Feynman integrals contributing
to the respective order needed. In this paper we describe a method, which allows to perform the inverse 
Mellin 
transform for massless and single--mass problems in these higher--order calculations. The method is instrumental 
in 
cases, where Mellin space representations cannot easily be derived. The method can also be applied in the 
presence
of more scales, leading to more involved iterated alphabets, however. Also the non--first--order--factorizing
equations become more involved due to real--valued parameters, additional cuts, etc.  

For massive operator matrix elements (OMEs), massless off--shell operator matrix elements or Wilson 
coefficients, a central variable $t$ can be identified, in which the differential equations of the 
respective master integrals are formulated. In the case of the OMEs this variable emerges through the 
resummation of the local composite operators into linear propagators and in the case of the Wilson 
coefficients it is the ratio $t = 2 p.q/Q^2$ of two kinematic invariants. Here $q^2 = -Q^2$ denotes the 
virtuality of the process, $q = l-l'$, with the initial state lepton ($l$), final state lepton ($l'$), 
and nucleon momentum ($p$). The corresponding series are {\it formal} Taylor series with $t \in \mathbb{R}$
and can be interpreted as generating functions.

While the variable $t$ emerges naturally in the case of Wilson coefficients, it has to be considered an 
auxiliary variable in the case of OMEs. At the end of the calculation one would like to perform the 
principal transformation
\begin{eqnarray}
t \rightarrow \pm \frac{1}{x},
\label{eq:TRF}
\end{eqnarray}
where $x = Q^2/(2 p.q)$ denotes the first Bjorken variable. As we will outline below, special care is necessary
because of the occurrence of $\delta(1-x)$ and of $+$-distributions \cite{DISTR} in $x$--space and one 
finally would like to consider different regions in $x$. In the case of deep--inelastic scattering this is 
$x \in [0,1]$.

  In previous calculations we have already made use of generating
  functions in $t$. However, in those cases we performed a formal Taylor
  expansion in which the $N$th Mellin moment arises as the coefficient
  of the expansion term $t^N$. In many cases it is possible to obtain
  the Mellin space result analytically 
\cite{Ablinger:2010ty,
HEAVY,
Ablinger:2014uka,
Ablinger:2014nga,
Ablinger:2014vwa,
Ablinger:2017tan,
Blumlein:2017wxd,
Blumlein:2021ryt,
Blumlein:2021enk,
Blumlein:2022gpp,
Ablinger:2019etw,
TWOMASS,Ablinger:2017xml,Ablinger:2019gpu}. 
One possibility is to
  calculate a large number of moments for the master integrals, assemble
  them into moments for the physical quantity that is being calculated,
  guess recurrences for them 
 \cite{GUESS1}
and finally solve those recurrences
  using the algorithms of the package \texttt{Sigma} 
\cite{SIG1,SIG2}.
  Calculating moments for the master integrals is often mathematically
  easier than computing them analytically.\footnote{In this way, we could compute the three--loop anomalous dimension 
$(\Delta) 
\gamma_{qg}^{(2)}(N)$ in a
massive environment, despite the fact that the master integrals contain elliptic structures, in Refs.~\cite{Ablinger:2017tan}.
As expected the elliptic structures cancel up to the $1/\ep$ terms in the final result, which is not evident 
by looking at the solutions of individual master integrals. Here $\ep = D - 4$ denotes the dimensional parameter.} 
For certain
  calculations we were able to push the number of moments that we could
  generate to $O(15000)$ \cite{ABLINGER1}. 

In the present paper we advocate for a complementary method. As a starting point the master integrals all have to 
be solved in analytic form in terms of the auxiliary variable $t$, 
including also eventual non first--order factorizing cases, which leads to iterated non--iterative integrals 
\cite{INII} in general. 
The expressions in the variable $t$ still have to be considered as a mathematical representation close to 
Mellin $N$--space. We will then construct the representations of the integrals in 
the Bjorken variable $x$ by analytic continuation. This method has already been applied in one of our recent calculations 
\cite{Ablinger:2022wbb} in the case of iterated integrals. In the present paper we will not treat OMEs in the 
two--mass case although the same method applies. 
Many of the contributions leading to iterated integrals have been calculated, 
cf.~\cite{TWOMASS,Ablinger:2017xml,Ablinger:2019gpu}. 
These integrals depend also on the real--valued mass ratio $m_c^2/m_b^2$ of the charm and bottom quark mass and the 
alphabet is square--root valued. 

The massive and massless OMEs are obtained from scattering amplitudes after performing the light--cone expansion
\cite{LCE}. Physically they are defined at integer values of the Mellin variable $N$ only. 
The set of all Mellin moments encodes the complete analytic information, cf.~\cite{Gross:1973ju}.
The corresponding $x$--space expressions, e.g. \cite{SPLIT}, are related via a Mellin transform
\begin{eqnarray}
\Mvec[f(x)](N) = \int_0^1 dx x^{N-1} f(x)
\label{eq:MEL}
\end{eqnarray}
to the former ones and have to be considered rather a derived quantity in general.\footnote{Curiously, in the 
massless case,
the corresponding lowest order functions were known about 50 years earlier before 
Mellin space representations have been considered \cite{Gross:1973ju}, which founded the method of 
equivalent photons, \cite{LALI}.}  
The inverse Mellin transform is given by
\begin{eqnarray}
f(x) = \frac{1}{2\pi i} \int_{c-i\infty}^{c+i\infty} ds \, x^{-s}
   \Mvec[f(x)](s),
\end{eqnarray}
where the integration contour surrounds all singularities of $\Mvec[f(x)](s)$ in the complex plane.
It will be shown below that the functions in $x$--space may have definitions on subsets or supersets of the interval 
$x \in [0,1]$ only, cf.~\cite{Ablinger:2017xml,Ablinger:2019gpu} at intermediate steps, and require
(various) distribution valued regularizations.

In this paper we apply the method outlined above
to integrals contributing to the massless and massive OMEs and massless 
Wilson coefficients to three--loop order and illustrate it by characteristic examples for the different function 
spaces. The calculation of these building blocks is of central importance for single--scale hard 
scattering cross 
sections in $pp, ep$ and $e^+e^-$ processes to three--loop order in QCD and QED. These results form also the basis
of precision measurements of the strong coupling constant $\alpha_s(M_Z^2)$ \cite{ALPHAS}, the value of the charm quark 
mass \cite{Alekhin:2012vu}, and precision determinations of the twist--2 parton distribution functions 
\cite{Accardi:2016ndt} at colliders such as HERA \cite{Blumlein:1987xk}, the LHC, and facilities 
planned for the future, such as EIC \cite{Boer:2011fh}, LHeC \cite{LHeCStudyGroup:2012zhm}, and the FCC 
\cite{FCC:2018vvp}. 

The paper is organized as follows. In Section~\ref{sec:2} we discuss the basic method for the inverse Mellin 
transform. 
In Section~\ref{sec:3} we show how to use our proposed method 
on different classes of iterated integrals, such as harmonic polylogarithms \cite{Remiddi:1999ew},
generalized harmonic polylogarithms \cite{Borwein:1999js,Moch:2001zr,Ablinger:2013cf}, 
cyclotomic harmonic polylogarithms 
\cite{Ablinger:2011te}, and iterated integrals containing square root valued letters \cite{Ablinger:2014bra}. 
In Section~\ref{sec:4} we investigate the case where also iterated 
non--iterative integrals are present, \cite{INII,Ablinger:2017bjx}.
In Section~\ref{sec:5} we comment on ways of efficient numerical 
representations of the results in $x$--space and Section~\ref{sec:6} contains the conclusions. 
Some technical aspects are given in the Appendices.
\section{The method} 
\label{sec:2}

\vspace*{1mm} 
\noindent
The Feynman rules for the local operator insertions are given in \cite{Bierenbaum:2009mv,Blumlein:2021ryt}
up to three--loop order. The operator matrix elements (OMEs) are proportional to $(\Delta.p)^N$, with $p$ 
the through flowing momentum, $\Delta$ a light--like vector and $N$ the Mellin index, which is given by an 
even or odd integer, depending on the physical problem. Furthermore, the crossing--relations 
\cite{Politzer:1974fr,Blumlein:1996vs} determine the range of the values $N \geq N_0,~~N,N_0 \in 
\mathbb{N} \backslash \{0\}$. One may resum all operator insertions by introducing an auxiliary parameter $t$ 
in terms of a formal Taylor series. For the simplest operator insertion one e.g. finds
\cite{Ablinger:2014yaa}
\begin{eqnarray}
\label{eq:resum}
\sum_{k=0}^\infty t^k (\Delta.p)^k = \frac{1}{1- t \Delta.p},~~~t \in \mathbb{R}.
\end{eqnarray}
The more involved operator insertions result in related structures, 
always leading to products of effective propagators as given in Eq.~(\ref{eq:resum}).
Respecting the crossing relations one has, more generally,
\begin{eqnarray}
\label{eq:resum1}
\sum_{k=0}^\infty t^k (\Delta.p)^k \frac{1}{2} [1 \pm (-1)^k] = 
\frac{1}{2} \left[
\frac{1}{1-t \Delta.p} \pm \frac{1}{1+ t \Delta.p} 
\right].
\end{eqnarray}
This representation has the advantage that the information on the operators is now fully contained in 
propagators and one may use the integration-by-parts (IBP) relations \cite{IBP} without specifying the 
different operator structures for each value of $N$, which grow rapidly in size for growing $N$.

The complete OMEs or the Wilson coefficients have a definite crossing behaviour, i.e.
\begin{eqnarray}
A^+(N) = \tfrac{1}{2} [1 + (-1)^N] A(N),
~~~~~\text{or}~~~~~
B^-(N) = \tfrac{1}{2} [1 - (-1)^N] B(N), 
\end{eqnarray} 
with
\begin{eqnarray} 
A(N) = \Mvec[A(x)](N),~~~~
B(N) = \Mvec[B(x)](N),
\end{eqnarray} 
and either only even or odd moments contribute. In $t$--space one obtains 
\begin{eqnarray} 
\tilde{A}(t) &=& \sum_{N=1}^\infty t^N A^+(N) = \int_0^1 dx \frac{t^2 x}{1 - t^2 
x^2} {A}(x),
~~\text{or}~~
\\
\tilde{B}(t) &=& \sum_{N=1}^\infty t^N B^-(N) = \int_0^1 dx \frac{t}{1 - t^2 x^2} {B}(x),
\end{eqnarray}
where
\begin{eqnarray}
\tilde{A}(t) = \tilde{A}(-t),~~~~~~~~~ 
\tilde{B}(t) = -\tilde{B}(-t).
\end{eqnarray}
Structures of the kind of Eq.~(\ref{eq:resum1}) also emerge in normal Feynman diagram calculations, such as for 
sub--system scattering processes or Wilson coefficient functions \cite{Vermaseren:2005qc,Blumlein:2022gpp}. Here 
the role of the 
parameter $t$ is 
taken by the fraction
\begin{eqnarray}
t = \frac{2 p.q}{Q^2}.
\end{eqnarray}

Let us get back to Eq.~(\ref{eq:resum1}), where the light--like vector $\Delta$ has been 
introduced.
By deriving the OMEs in the light--cone expansion \cite{LCE}, cf.~\cite{Politzer:1974fr,Blumlein:1996vs}, in 
Fourier--space their $N$th moment scales also with $1/x^n$, cf.~e.g.~Eqs.~(52,53) of Ref.~\cite{Blumlein:1996vs}.
Therefore, the situation is the same as in the case of the Wilson coefficients.
The resummed OMEs behave mathematically very similar to the forward Compton amplitude 
$T_{\mu\nu}$ \cite{Vermaseren:2005qc,Blumlein:2022gpp}
and one may formally use the relation
\begin{eqnarray}
\label{eq:opt}
W_{\mu\nu} =  \frac{1}{\pi} {\sf Im} \, T_{\mu\nu},~~~\forall t \in \mathbb{R}.
\end{eqnarray}
Here $W_{\mu\nu}$ is the hadronic tensor. For the present application $W_{\mu\nu}$ is the final function
depending on the Bjorken variable $x$, while $T_{\mu\nu}$ contains the variable $t$. 
The imaginary part in (\ref{eq:opt}) results from the monodromy of the 
iterated, or iterated non--iterative integrals \cite{INII} 
around $t = 1$, $t = -1$, and complex valued contributions of other kind by setting
\begin{eqnarray}
\label{eq:TRA1}
t = \pm \frac{1}{x}.
\end{eqnarray}
This way the result in $x$--space can be obtained.
Because of even and odd moments being present in intermediate results, one has to consider also the case 
$t = - 1/x$, according to the cuts in the forward Compton amplitude, cf.~\cite{Politzer:1974fr,
Geyer:1977gv,Buras:1979yt,Reya:1979zk,Blumlein:2012bf}. In the case of iterated integrals,
the monodromy is described by the Drinfeld--Knizhnik--Zamolodchikov \cite{DKZ} equations and related 
equations. Special care has to be taken in the case of distribution--valued contributions in 
$x$--space, cf.~Section~\ref{sec:21}.

In Eq.~(\ref{eq:opt}) only the main cut is considered, since all hadronic cut--(final) states are
summed over. This relation applies also to the individual Feynman diagrams and the associated scalar
integrals. Equivalently, one may
consider the associated (subtracted) dispersion relations, cf.~\cite{Blumlein:1996vs}, also known 
as Kramers--Kronig relation \cite{KK} or K\"allen--Lehmann representation \cite{KL}.

In the following, we will elaborate on the extraction of the $x$--space representation by 
analytic continuation of the generating function expressed in $t$.
Let us consider a function ${F}(N)$ which has the representation
\begin{eqnarray}
{F}(N) = \int_0^1 dx x^{N-1}[f(x) + (-1)^{N-1} g(x)],
\end{eqnarray}
with $f(x) = g(x) = 0,~~\text{for}~~x \in \mathbb{R}, x < 0, x > 1$. 
Terms of this kind appear e.g. in the flavor non--singlet 
anomalous dimensions \cite{Moch:2004pa,Blumlein:2021enk,Blumlein:2011mi}.

Its $t$--representation is then given by
\begin{eqnarray}
\tilde{F}(t) = \sum_{N=1}^\infty t^N {F}(N) = \int_0^1 
dx'\left[
\frac{t f(x')}{1-tx'} + \frac{t g(x')}{1+tx'}\right].
\end{eqnarray}
For the physical variable $x \in [0,1]$ one finds
\begin{eqnarray}
\label{eq:disc1}
\tilde{F}\left(t = \frac{1}{x}\right) &=& \int_0^1 dx' \left[
\frac{f(x')}{x-x'} + \frac{g(x')}{x+x'}\right],
\\
\label{eq:disc2}
\tilde{F}\left(t = -\frac{1}{x}\right) &=& \int_0^1 dx' \left[
-\frac{f(x')}{x+x'} + \frac{g(x')}{x'-x}\right].
\end{eqnarray}
We can use the Sochocki formulae \cite{DISTR,SOCH}\footnote{These relations can also be derived
by using residue theory, cf.~\cite{PRIWALOW}, see Appendix~\ref{sec:A}.}
\begin{eqnarray}
\lim_{\delta \rightarrow 0^+} \frac{1}{\xi \pm i \delta} = {\cal P} \frac{1}{\xi} \mp i \pi \delta(\xi)
\end{eqnarray}
with ${\cal P}$ Cauchy's valeur principale \cite{CAUCHY}, to replace 
the denominators in (\ref{eq:disc1}, \ref{eq:disc2}) with $\xi = x 
\pm x'$ and obtain
\begin{eqnarray}
-\frac{1}{2 \pi i}{\sf Disc}_x \tilde{F}\left(\frac{1}{x}\right) &=& 
\lim_{\delta \rightarrow 0^+} 
\frac{1}{\pi}{\sf Im} \tilde{F}\left(\frac{1}{x - {\rm i} \delta}\right) = 
\int_0^1 dx' f(x') \delta(x - x') =  
f(x),
\label{eq:DISC1}
\\
\frac{1}{2 \pi i} {\sf Disc}_x \tilde{F}\left(-\frac{1}{x}\right) &=& 
\lim_{\delta \rightarrow 0^+} 
 \frac{1}{\pi} {\sf Im} \tilde{F}\left(-\frac{1}{x + {\rm i}\delta}\right) =
\int_0^1 dx' g(x') \delta(x - x') =  g(x).
\label{eq:DISC2}
\end{eqnarray}
One therefore may reconstruct
\begin{eqnarray}
\label{eq:fg}
f(x) + (-1)^{N-1} g(x) = \frac{1}{2 \pi i}\left[
-{\sf Disc}_x \tilde{F}\left(\frac{1}{x}\right)
+ (-1)^{N-1} {\sf Disc}_x \tilde{F}\left(-\frac{1}{x}\right) \right].
\end{eqnarray}
One realizes that the branch of the solution that scales proportional to 
$(-1)^N$ introduces a monodromy at the point $t=-1$, which has to be 
accounted for. 
Similarly, one may consider branches which scale more generally as 
$r^N, \ r \in \mathbb{R}$, introducing a monodromy at a $t=1/r$, which has 
to be handled accordingly and will lead to $x$--space representations 
with support different from $x \in ]0,1[$. 
For single iterated integrals and Feynman diagrams this has been already 
observed in the case of the massive pure singlet and 
two--mass OMEs.
However, in the physical amplitude these contribution outside of the physical 
region canceled and one was left with the usual support.

We illustrate our method with the following example,
\begin{eqnarray}
\label{eq:EX11}
\tilde{F}(t) &=& \HA_{0,1,-1}(t) + 2 \HA_{0,0,-1}(t).
\end{eqnarray}
The Mellin space expression, corresponding to the coefficient of $t^N$, reads
\begin{eqnarray}
\Mvec[F(x)](N) &=& \frac{(-1)^{N-1}}{N^3} - \frac{S_{-1}(N)}{N^2}
\end{eqnarray}
describing the expansion coefficients of (\ref{eq:EX11})
\begin{eqnarray}
\tilde{F}(t) =
2 t+\frac{7 t^3}{54}+\frac{t^4}{48}+\frac{59 t^5}{1500}
+\frac{t^6}{80}+\frac{379 t^7}{20580}+\frac{107 t^8}{13440} 
+ O(t^9).
\end{eqnarray}
Here $S_{\vec{a}}(N)$ denotes the harmonic sums 
\cite{Vermaseren:1998uu,Blumlein:1998if}
\begin{eqnarray}
S_{b,\vec{a}}(N) = \sum_{k=1}^N \frac{({\rm sign}(b))^k}{k^{|b|}} 
S_{\vec{a}}(k),~~~S_\emptyset = 1,~~b, a_i \in \mathbb{Z} \backslash \{0\},~~N \in 
\mathbb{N} 
\backslash \{0\}. 
\end{eqnarray}
One obtains the following functions in (\ref{eq:fg})
\begin{eqnarray}
f(x) &=& - \ln(2) \HA_0(x)
\\
g(x) &=& - \ln(2) \HA_0(x) -\frac{1}{2}\zeta_2 + \frac{1}{2} \HA_0^2(x) + \HA_{0,-1}(x).
\end{eqnarray}

Let us likewise consider the functions with definite crossing relations
\begin{eqnarray}
{F}(x) = [1 + (-1)^{N-1}] s(x) + [1-(-1)^{N-1}] a(x),
\end{eqnarray}
and resum its Mellin transform into $\tilde{F}(t)$,
\begin{eqnarray}
\tilde{F}(t) &=& \sum_{N=1}^\infty t^N \Mvec[F(x)](N) 
\nonumber\\ 
&=& \int_0^1 dx' 
x'^{N-1} \Biggl[ 
\frac{2 t s(x')}{1-t^2 x'^2}
+ \frac{2 t^2 x' a(x')}{1-t^2 x'^2}
\Biggr],
\nonumber\\ &=&
\int_0^1 dx' x'^{N-1} \Biggl[
\frac{t}{1-t x'} (s(x') + a(x'))
+ \frac{t}{1+t x'} (s(x') - a(x'))\Biggr].
\end{eqnarray}
One obtains the combinations
\begin{eqnarray}
\label{eq:sa}
s(x) +  a(x) = - \frac{1}{2 \pi i}{\sf Disc}_x F\left(\frac{1}{x}\right),~~~~~~~~~
s(x) -  a(x) =   \frac{1}{2 \pi i}{\sf Disc}_x F\left(-\frac{1}{x}\right).
\end{eqnarray}
In cases which are free of the factor $(-1)^{N-1}$ in $x$--space it is 
sufficient to consider $\tilde{F}(t = 1/x)$ since the monodromy around $t = -1$ does not play a role. 
Most of the cases discussed below receive, however, contributions form both terms. 
On the other hand, it is evident that in the case
that either $s(x)$ or $a(x)$ vanish, one of the equations (\ref{eq:sa}) is sufficient to determine the
respective distribution. 

The strategy to apply Eq.~(\ref{eq:fg}) is now to first analytically 
calculate the master integral in terms of iterated non--iterative integrals in the variable $t$, describing the 
resummed Mellin--space representation. 
This is done by solving the corresponding 
systems of linear ordinary differential equations over arbitrary bases of master integrals, as has been described 
in 
Ref.~\cite{Ablinger:2018zwz}. The iterated non--iterative integrals 
are then found by solving the homogeneous solutions in terms 
of higher transcendental functions and the application of Euler--Lagrange 
\cite{EULER1} variation of the constant.
This is followed by the transformation $t \to \pm 1/x$ and applying
  \ref{eq:fg}, leading to another analytic iterated non-iterative integral.
  These integrals now depend on the Bjorken variable $x$, which is
  identical to the momentum fraction variable $z$ in collinear
  factorization \cite{FEYNM} for twist--2 operators and forward scattering that
  we deal with in the present paper.
All expressions in $t$--space are understood as generating functions and the resummation
(\ref{eq:resum1}) traces the $N$--space solution as the $N$th expansion coefficient. 

Concrete master integrals were derived in different projects \cite{Ablinger:2014nga,
Ablinger:2015tua,Blumlein:2021ryt,Blumlein:2021enk,Blumlein:2022gpp,Ablinger:2022wbb} 
which were obtained by using e.g. the packages {\tt Reduze~2} \cite{REDUZE} and {\tt Crusher} 
\cite{CRUSHER}. We note that the Mellin $N$ result for each contributing power in $N$ can 
be directly obtained by expanding in $t$. 
We will demonstrate our new method of directly obtaining the $x$--space expression from the 
generating function in $t$ on different function classes which arose in the 
aforementioned projects in Sections~\ref{sec:3} and~\ref{sec:4}.

In calculating massless and massive OMEs different alphabets forming iterated and 
iterated non--iterative integrals were 
revealed. The words formed out of these alphabets encode the whole information of the respective Quantum Field
Theory\footnote{One may call these alphabets also the genetic code of the micro cosmos, cf. \cite{JB17}.}, 
like other alphabets provide the basic building blocks for languages and other structures \cite{STRUCT}. 
The simplest one is formed by the harmonic polylogarithms (HPLs) \cite{Remiddi:1999ew} and its subsets,
the classical \cite{LEWIN} and Nielsen polylogarithms \cite{NIELSEN}. These are followed by generalized 
harmonic polylogarithms \cite{Borwein:1999js,Moch:2001zr,Ablinger:2013cf}, cyclotomic polylogarithms 
\cite{Ablinger:2011te}, and specific 
root--valued alphabets obtained in Mellin inversions of finite binomial and inverse central binomial sums 
\cite{Ablinger:2014bra,Ablinger:2022wbb}. All these alphabets lead to iterated integrals, for which shuffle 
algebras \cite{ALG} lead to a reduction of the respective representation. 

In massive problems at three--loop order also $_2F_1$--letters occur, cf. e.g.~\cite{Ablinger:2017bjx}, 
which are no iterated integrals anymore. They can be dealt with in terms of iterated non--iterative 
integrals, however. Going even to higher orders, more and more of these structures will occur. They are 
characterized e.g. as solutions of differential equations, which do not factorize at first order.
The $_2F_1$--letters are related to complete elliptic integrals \cite{TRICOMI} of specific (irrational) functions 
in $t$ and 
to modular forms \cite{MODULAR}. We also note that among 
square--root letters one may have those, leading to 
incomplete 
elliptic integrals, cf.~\cite{Blumlein:2020jrf}. These cases, however, 
are iterated integrals.
We remark, that transformations like (\ref{eq:TRA1}) also connect splitting functions 
with argument $x \in [0,1]$ 
to fragmentation functions with $x \in [1,\infty)$, cf.~\cite{Blumlein:2000wh}.
In Sections~\ref{sec:3} and \ref{sec:4} we will demonstrate the present method for the different classes 
of functions mentioned above and illustrate it by a series of examples.

In the next section we describe the separation of the different distribution--valued contributions in 
$x$--space directly from the $t$--space representation in Section~\ref{sec:21}, and the property of conjugation, 
which relates different master integrals and can be used to decrease the number of master integrals which have to 
be calculated, in Section~\ref{sec:23}.
\subsection{Distributions in \boldmath $x$ space}
\label{sec:21}

\vspace*{1mm}
\noindent
In inclusive physical (single--scale) processes there occur two distribution--valued contributions,
\begin{eqnarray}
\label{eq:distr}
\delta(1-x),~~~~~~~\left(\frac{\ln^k(1-x)}{1-x}\right)_+,~k 
\in \mathbb{N},~~~\text{with}~~~\ln(1-x) = - \HA_1(x),
\end{eqnarray}
where $\HA_a(x)$ denotes a harmonic polylogarithm \cite{Remiddi:1999ew}. They describe the soft region
$x \rightarrow 1$ or $N \rightarrow \infty$.
Both distributions emerge from the behaviour of the generating function at $t=1$. Ideally one would like to separate
these contributions in $t$--space already, since their $x$--space structure is known, such 
that finally only the regular part needs to be calculated in $x$ space. The Mellin transform of the 
distributions read
\begin{eqnarray}
\Mvec[\delta(1-x)](N) &=& 1,
\\
\Mvec\left[\left(\frac{f_a(x)}{1-x}\right)_+\right](N) &=& \int_0^1 dx \frac{x^{N-1} - 1}{1-x} f_a(x), 
\end{eqnarray}
which is the option {\tt PlusFunctionDefinition $\rightarrow$ 1} of the package {\tt HarmonicSums} 
\cite{HARMSU,Blumlein:2009ta,Vermaseren:1998uu,Blumlein:1998if,Remiddi:1999ew,Ablinger:2011te,
Ablinger:2013cf,Ablinger:2014bra,ALG,Ablinger:2021fnc}. For the separation of the distribution 
we will consider $f_a(x) = \HA_1^k(x),~~k \in \mathbb{N}$, for definiteness. Details of the 
decomposition in the $\delta, +$ and regular contribution are given in Appendix~\ref{sec:A}.

One expands the analytic solution $G(t)$ around $t=1$ as 
\begin{eqnarray}
G(t) \simeq \frac{1}{1-t} a_0 + \sum_{k = 1}^\infty a_k \frac{\HA_1^k(t)}{t-1} + \hat{G}_{\rm reg}(t),
\end{eqnarray}
with $\hat{G}_\text{reg}(t) = O\left((t-1)^0\right)$ and  $\hat{G}_{\rm reg}(t)$ does not result 
in a distribution  in $x$--space. By this one obtains the 
leading terms contributing to the distributions. To obtain the complete distributions in $x$--space one subtracts from 
$G(t)$ the following distribution--generating
terms, with the coefficients $a_k$ (\ref{eq:dist1}--\ref{eq:dist5}), etc.,
leaving ${G}_{\rm reg}(t)$, a modified form of $\hat{G}_{\rm reg}(t)$.

In this way, one identifies the leading terms in the $t$--representation. 
The distribution--valued contributions are obtained by the following replacements
\begin{eqnarray}
\label{eq:dist1}
\delta(1-x)
&\leftarrow& 
\frac{t}{1-t},  
\\
\left[\frac{1}{1-x}\right]_+
&\leftarrow& \frac{t}{t-1} \HA_1(t), 
\\
\left[\frac{\HA_1(x)}{1-x}\right]_+ 
&\leftarrow& 
\frac{t}{t-1} \left[ \frac{1}{2} \HA^2_1(t) + \HA_{0,1}(t)
\right],
\\
\left[\frac{\HA_1^2(x)}{1-x}\right]_+
&\leftarrow&
\frac{t}{t-1} \left[ 
\frac{1}{3} \HA_1^3(t) + 2 \HA_1(t) \HA_{0,1}(t) + 2 \HA_{0,0,1}(t) - 2 \HA_{0,1,1}(t)
\right],
\\
\left[\frac{\HA_1^3(x)}{1-x}\right]_+
&\leftarrow&
\frac{t}{t-1} \Biggl[
\frac{1}{4} \HA_1^4(t)
        +3 \HA_1^2(t) \HA_{0,1}(t)
        +6 \HA_ 1(t) \HA_{0,0,1}(t)
        -6 \HA_ 1(t) \HA_{0,1,1}(t)
\nonumber\\ && 
        +6 \HA_{0,0,0,1}(t) 
        -6 \HA_{0,0,1,1}(t)
        +6 \HA_{0,1,1,1}(t)
\Biggr]
\\
\left[\frac{\HA_1^4(x)}{1-x}\right]_+
&\leftarrow& \frac{t}{t-1} \Biggl[
\frac{1}{5} \HA_1^5(t)
+4 \HA_1^3(t) \HA_{0,1}(t)
+12 \HA_1^2(t) \HA_{0,0,1}(t)
-12 \HA_1^2(t) \HA_{0,1,1}(t)
\nonumber\\ && 
+24 \HA_1(t) \HA_{0,0,0,1}(t)
-24 \HA_1(t) \HA_{0,0,1,1}(t)
+24 \HA_1(t) \HA_{0,1,1,1}(t)
+24 \HA_{0,0,0,0,1}(t)
\nonumber\\ && 
-24 \HA_{0,0,0,1,1}(t)
+24 \HA_{0,0,1,1,1}(t)
-24 \HA_{0,1,1,1,1}(t)
\Biggr]
\\
\label{eq:dist5}
\left[\frac{\HA_1^5(x)}{1-x}\right]_+
&\leftarrow&  
\frac{t}{t-1} \Biggl[
    \frac{1}{6} \HA_1^6(t)
    +5 \HA_1^4(t)  \HA_{0,1}(t)
    +20 \HA_1^3(t) \HA_{0,0,1}(t)
    -20 \HA_1^3(t) \HA_{0,1,1}(t)
\nonumber\\ &&
    +60 \HA_1^2(t) \HA_{0,0,0,1}(t)
    -60 \HA_1^2(t) \HA_{0,0,1,1}(t)
    +60 \HA_1^2(t) \HA_{0,1,1,1}(t)
\nonumber\\ &&
    +120 \HA_1(t)  \HA_{0,0,0,0,1}(t)
    -120 \HA_1(t)  \HA_{0,0,0,1,1}(t)
    +120 \HA_1(t)  \HA_{0,0,1,1,1}(t)
\nonumber\\ &&
    -120 \HA_1(t)  \HA_{0,1,1,1,1}(t)
    +120 \HA_{0,0,0,0,0,1}(t)
    -120 \HA_{0,0,0,0,1,1}(t)
\nonumber\\ &&
    +120 \HA_{0,0,0,1,1,1}(t)
    -120 \HA_{0,0,1,1,1,1}(t)
    +120 \HA_{0,1,1,1,1,1}(t)
  \Biggr],
~~~\text{etc.}
\end{eqnarray}
In the substitution one shall start from the largest power $k$ in Eq.~(\ref{eq:distr}). One notices that the 
coefficients of the formal Taylor series of these expressions are the same as the values of the Mellin moments
of the distributions at the l.h.s.\footnote{We remark that {\tt Mathematica} and {\tt HarmonicSums} have partly 
different implementations of cuts.}

\subsection{Conjugation}
\label{sec:23}

\vspace*{1mm}
\noindent
In the calculation of single--scale master integrals finally expressed in the variable $x$ in momentum fraction 
space, one observes, in quite a series of cases, the so-called conjugation relation. In 
Mellin $N$--space it reads, 
cf.~\cite{Vermaseren:1998uu}, 
\begin{eqnarray}
\hat{f}_2(N,\ep) \equiv \hat{f}_1^C(N,\ep) = - \sum_{k=1}^N (-1)^k \binom{N}{k} \hat{f}_1(k,\ep),
\end{eqnarray}
for the functions $\hat{f}_1(N,\ep)$ and $\hat{f}_2(N,\ep)$, at all orders in the dimensional 
parameter $\ep$.
One may phrase this relation in $x$--space directly with
\begin{eqnarray}
\hat{f}(N,\ep) = \Mvec[{f}(x,\ep)](N) \equiv \int_0^1 dx~x^{N-1}~{f}(x,\ep),
\end{eqnarray}
yielding
\begin{eqnarray}
{f}_2(x,\ep) =   -\frac{x}{1-x} f_1(1-x,\ep),~~~~\text{for}~~x \in [0,1[.
\label{eq:conj1}
\end{eqnarray}
The conjugation relation obeys
\begin{eqnarray}
[\hat{f}^C(N)]^C = \hat{f}(N),~~~~[{f}^C(x)]^C = \tilde{f}(x).
\end{eqnarray}
The most simple example is 
\begin{eqnarray}
S_1^C(N) = \frac{1}{N},
\end{eqnarray}
reading in $x$--space
\begin{eqnarray}
\left(- \frac{x}{1-x}\right)^C = 1. 
\end{eqnarray}
Some of the master integrals are even self--conjugate. It is useful to study 
a large number of moments of all master integrals first, to find those which are conjugate to others, since 
their direct calculation can be avoided by using Eq.~(\ref{eq:conj1}). This has been done also for the
massive OME $A_{Qg}^{(3)}$ \cite{Blumlein:2017wxd}.

\section{Iterated integrals} 
\label{sec:3}

\vspace*{1mm} 
\noindent
Iterated integrals $\GA({a_1}, ..., {a_k};t)$ are defined over an alphabet $\mathfrak{A}$
\begin{eqnarray}
\mathfrak{A} = \{f_1(t), ..., f_m(t)\}
\end{eqnarray}
of letters $f_k(t)$ which are analytic functions of $t$. They are given by
\begin{eqnarray}
\GA(b,\vec{a};t) = \int_0^t dx_1 f_b(x_1) \GA(\vec{a};x_1).
\label{eq:GA1}
\end{eqnarray}
If one of the letters $f_k(t)$ behaves like $c_k/t, c_k \in \mathbb{C} \backslash \{0\}$ the integral
$\int_0^x dt~f_k(t)$ needs a regularization given by
\begin{eqnarray}
\label{eq:REG}
\GA(k;x) := \int_\ep^x dt~f_k(t) + \HA_0(\ep),
\end{eqnarray}
which leads to regulators $\propto \ln^l(\ep)$ that have to cancel in the final expression. Examples are
\begin{eqnarray}
\GA(0;x) &:=& \int_\ep^x dt~\frac{1}{t} + \HA_0(\ep) = \HA_0(x), 
\\
\GA\left(\frac{\sqrt{1+x}}{x};x\right) &:=& \int_{\ep}^x \frac{dy}{y} \sqrt{1-y} + 
\HA_0(\ep) =
-2 + 2 \sqrt{1 - x} + 2 \ln(2) 
\nonumber\\ &&
+ \ln(1 - \sqrt{1 - x}) - \ln(1 + \sqrt{1 - x}).
\end{eqnarray}
These regularizations are necessary for the letter $1/t$ contributing to the harmonic polylogarithms and
to several other alphabets.

The iterated integrals obey the recurrent differential equation
\begin{eqnarray}
\frac{1}{f_b(t)} \frac{d}{dt} \GA(b,\vec{a};t) = \GA(\vec{a};t),
\end{eqnarray}
which can be iterated to yield a first--order--factorizing differential equation for $G(b,\vec{a};t)$ itself,
\begin{eqnarray}
\left[\frac{d}{dt} 
\frac{1}{f_{a_{k-1}}(t)} \frac{d}{dt} 
...
\frac{1}{f_{a_1}(t)} \frac{d}{dt} \right]
\GA(\vec{a};t) = f_{a_k}(t).
\end{eqnarray}
One may now perform the transformation $t \rightarrow 1/x$, which yields
\begin{eqnarray}
\left[- x^2 \frac{d}{dx} 
\frac{(-x^2)}{f_{a_{k-1}}\left(\frac{1}{x}\right)} \frac{d}{dx} 
~...~ 
\frac{(-x^2)}{f_{a_1}\left(\frac{1}{x}\right)}  \frac{d}{dx} \right]
\GA\left(\vec{a};\frac{1}{x}\right) = f_{a_k}\left(\frac{1}{x}\right).
\label{eq:GLx}
\end{eqnarray}
The boundary conditions for the solution of (\ref{eq:GLx}) are known by $G(\vec{a};t=1)$. From 
$\tilde{F}(t) = \GA\left(\vec{a};t\right)$ one obtains from (\ref{eq:GLx}) $\tilde{\tilde{F}}(x) = 
\GA\left(\vec{a};1/x\right)$ and 
\begin{eqnarray}
F(x) = \frac{1}{\pi} {\sf Im} \,\tilde{\tilde{F}}(x),
\end{eqnarray}
and similarly for $t \rightarrow -1/x$.
In this way, all the corresponding calculations for the iterated integrals can be performed.
In various applications we will derive also the differential equations for the respective $\GA$--functions 
of the variable $\pm 1/x$, to extract the imaginary part.

\subsection{Harmonic polylogarithms} 
\label{sec:31}

\vspace*{1mm} 
\noindent
Harmonic polylogarithms \cite{Remiddi:1999ew} 
are the simplest entities in single--scale higher--loop calculations in QCD and QED.
Advanced examples where they appear and are sufficient to express the
final results are the massless three-loop Wilson coefficients \cite{Vermaseren:2005qc,Blumlein:2022gpp}. 
The alphabet is given by
\begin{eqnarray}
\mathfrak{A}_{\rm HPL} = \left\{f_0(x) = \frac{1}{x}, f_1(x)= \frac{1}{1-x}, f_{-1}(x) = \frac{1}{1+x} \right\}.
\end{eqnarray}
The HPLs are defined by\footnote{The summary--index notation used e.g. in \cite{Remiddi:1999ew}, e.g. 
writing the index 2 for $\{0,1\}$, is not used here.}
\begin{eqnarray}
\HA_{b,\vec{a}}(x) = \int_0^x dy f_b(y) \HA_{\vec{a}}(y),~~f_c \in \mathfrak{A}_{\rm HPL},~~
\HA_{\underbrace{\mbox{\tiny{0,...,0}}}_{k}}(x) := \frac{1}{k!} \ln^k(x).
\end{eqnarray}
in the $\HA_{\vec{b}}(x)$--notation.
We consider the functions\footnote{The labels 0, 1, and -1 refer to the usual HPL letters.}
\begin{eqnarray}
\tilde{F}_1(t) &=& \HA_{0, 0,  1}(t),
\\
\tilde{F}_2(t) &=& \HA_{0, 1, -1, 0, 1}(t).
\end{eqnarray}
For the first function the transformations $t \rightarrow \pm 1/x$ yields
\begin{eqnarray}
F_1\left(t = \frac{1}{x}\right) &=& - 2 \zeta_2 \HA_0(x) + \frac{1}{6} \HA^3_0(x) + \HA_{0,0,1}(x) + 
\frac{i\pi}{2} 
\HA^2_{0}(x),
\\
\label{eq:x1}
F_1\left(t = -\frac{1}{x}\right) &=& \zeta_2 \HA_0(x) + \frac{1}{6} \HA_0^3(x) - \HA_{0,0,-1}(x),
\end{eqnarray}
and one obtains
\begin{eqnarray}
F_1(x) &=& \frac{1}{2} \HA_0^2,
\end{eqnarray}
Here, (\ref{eq:x1}) does not contribute. The Mellin transform of $F_1(x)$ is
\begin{eqnarray}
\Mvec[F_1(x)](N) &=& \frac{1}{N^3},
\end{eqnarray}
which describes the $t$--series expansion of $\tilde{F}_1(t)$,
\begin{eqnarray}
\tilde{F}_1(t) = \sum_{N=1}^\infty \frac{t^N}{N^3}.
\end{eqnarray}

Similarly, one obtains $F_2(x)$
\begin{eqnarray}
F_2(x) = F_{2a}(x) + (-1)^{N-1} F_{2b}(x),
\end{eqnarray}
with
\begin{eqnarray}
F_{2a}(x) &=& 
-4 \Li_4\left(\frac{1}{2}\right)
-\frac{1}{6} \ln^4(2)
+\ln^2(2) \zeta_2
+\frac{103}{40} \zeta_2^2
+ \HA_{0,-1,0,1}
-\frac{1}{24} \HA_0^4
-\frac{1}{2} \HA_0^2 \HA_{0,1}
- \HA_{0,-1} \HA_{0,1}
\nonumber\\ && 
+ \HA_0 [2 \HA_{0,0,1}
+ \HA_{0,0,-1}
+ \HA_{0,1,-1}]
-3 \HA_{0,0,0,1}
-3 \HA_{0,0,0,-1}
+2 \HA_{0,0,-1,1}
+\frac{1}{2} \ln(2) \zeta_2 \HA_0
\nonumber\\ && 
+\frac{1}{4} \zeta_2 \HA_0^2 
+\frac{1}{2} \zeta_2 \HA_{0,1} 
+\frac{3}{2} \zeta_3 \HA_0,
\\
F_{2b}(x) &=& - \Biggl[
        -\frac{1}{2} \ln(2) \HA_0
        -\frac{1}{4} \HA_0^2
        +\frac{1}{2} \HA_{0,-1}
        -\frac{1}{4} \zeta_2
\Biggr] \zeta_2,
\end{eqnarray}
where we set $\HA_{\vec{a}}(x) \equiv \HA_{\vec{a}}$.
The Mellin transform of $F_2(x)$ is given by
\begin{eqnarray}
\Mvec[F_2(x)](N) = -\frac{1}{N^5}
+\left(
        \frac{(-1)^N}{N^3}
        -\frac{S_{-1}}{N^2}
\right) S_{-2}
+\frac{S_{-2,-1}}{N^2},
\label{eqF2x}
\end{eqnarray}
with the convention $S_{\vec{a}}(N) \equiv S_{\vec{a}}$. The first terms of the series of 
$\tilde{F}_2(t)$ read
\begin{eqnarray}
\tilde{F}_2(t) = \frac{t^3}{18}+\frac{t^4}{64}+\frac{67 t^5}{3600}+\frac{11 t^6}{1296}
+\frac{9619 t^7}{1058400}+\frac{7117 t^8}{1382400} + O(t^9),
\end{eqnarray}
in accordance with (\ref{eqF2x}). The constants are all multiple zeta values \cite{Blumlein:2009cf}. 
In this case the package {\tt HarmonicSums} provides the corresponding transformation. 
\subsection{Cyclotomic harmonic polylogarithms} 
\label{sec:32}
\vspace*{1mm} 
\noindent
The first letters of the cyclotomic alphabet read \cite{Ablinger:2011te}
\begin{eqnarray}
\mathfrak{A}_{\rm cycl} &=& \left\{\frac{1}{x}\right\} \cup \Biggl\{
\frac{1}{1-x},
\frac{1}{1+x},
\frac{1}{1 + x + x^2},
\frac{x}{1 + x + x^2}
\frac{1}{1  + x^2},
\frac{x}{1  + x^2},
\frac{1}{1 - x + x^2},
\nonumber\\ &&
\frac{x}{1 - x + x^2}, ...
\Biggr\}.
\end{eqnarray}
Here the highest numerator power of $x$ is given by Euler's totient function of 
the polynomial number, the
denominators are formed by the cyclotomic polynomials\footnote{One may also study iterated integrals given
by quadratic forms, cf.~\cite{Ablinger:2021fnc}.} and $\mathfrak{A}_{\rm HPL} \subset \mathfrak{A}_{\rm cycl}$ 
holds. The cyclotomic polylogarithms are defined by 
\begin{eqnarray}
\HA_{\{c_1,d_1\},\{a_{i_1},b_{i_1}\}, \ldots, \{a_{i_k},b_{i_k}\}}(x)
=
\int_0^x dy f_{\{c_1,d_1\}}(y)
\HA_{\{a_{i_1},b_{i_1}\}, \ldots \{a_{i_k},b_{i_k}\}}(y),
\end{eqnarray}
where $c_1, a_{i,k}$ label the cyclotomic polynomial and $d_1, b_{i_k}$ denote the degree of the numerator powers. 
Here and in the following we are referring to $\GA$--functions, always related to the alphabet 
discussed in the respective section. 

In physics applications cyclotomic polylogarithms were generated by the
third, fourth, and sixth cyclotomic polynomial, see e.g. 
\cite{CYCLPH,Ablinger:2014yaa,Ablinger:2015tua,Ablinger:2018zwz}. They also appear while calculating OMEs and 
Wilson 
coefficients for even/odd moments separately \cite{Blumlein:2021enk,Blumlein:2021ryt,Blumlein:2022gpp}.

We consider the following example 
\begin{eqnarray}
\tilde{F}_3(t) &=& 
\frac{1}{3 (1-t) t^{1/3}}
     \GA\left[\frac{\xi^{1/3}}{1-\xi}; t\right]
\\
&=&
\frac{1}{1-t} \left(
      -1
      +\frac{t^{-1/3}}{3} \left(
        \HA_{1}(t^{1/3})
        + 2 \HA_{\{3,0\}}(t^{1/3})
        + \HA_{\{3,1\}}(t^{1/3})
      \right)
    \right).
\end{eqnarray}
The first terms of its series expansion around $t=0$ read
\begin{eqnarray}
\tilde{F}_3(t) = 
\frac{t}{4}+\frac{11 t^2}{28}+\frac{69 t^3}{140}+\frac{1037 t^4}{1820}+\frac{4603 t^5}{7280}
+\frac{94737 t^6}{138320}+\frac{1111267 t^7}{1521520}+\frac{5860639 t^8}{7607600}
+O(t^9).
\label{eq:cyclt}
\end{eqnarray}
As the next step, one has to separate the distribution--valued terms first by expanding around $t=1$.
One finds the distributions
\begin{eqnarray}
a_1 \left[\frac{1}{1-x}\right]_+ 
+ a_0 \delta(1-x);~~~ a_1 = 
-\frac{1}{3},~~~ 
a_0 =  \frac{1}{18}\left[\sqrt{3} \pi + 9(-2 + \ln(3))\right]
\end{eqnarray}
and has to subtract $t/(t-1) [-a_0 + a_1 \HA_1(t)]$, before converting to the regular term in $x$ 
space.
Finally one obtains
\begin{eqnarray}
F_3(x) = -\frac{1}{3} \left[\frac{1}{1-x}\right]_+ + \frac{1}{18}\left[\sqrt{3} \pi + 9(-2 + \ln(3))\right]
\delta(1-x) + \frac{1 - x^{4/3}}{3 (1-x)}
\end{eqnarray}
and for the Mellin transform the following cyclotomic sum
\begin{eqnarray}
\Mvec[F_3(x)](N) &=& \sum_{k=1}^N \frac{1}{1+3k},
\end{eqnarray}
describing the pattern in (\ref{eq:cyclt}).
The transformation implies the contribution of cyclotomic constants, like $\pi, \ln(3)$ etc.,
cf.~\cite{Ablinger:2011te}.
\subsection{Generalized harmonic polylogarithms} 
\label{sec:33}

\vspace*{1mm} 
\noindent
The alphabet for this class of integrals is given by \cite{Ablinger:2013cf}
\begin{eqnarray}
\mathfrak{A}_{\rm gHPL} &=& \left\{\frac{1}{x-a}\right\},~~a \in \mathbb{C}.
\end{eqnarray}
For single--scale OMEs one has $a \in \mathbb{Z}$ or $\mathbb{Q}$. 
Alternatively, for $a,b_i \in \mathbb{R}$ we can also use the
  notation
  \begin{align}
    H_{a,\vec{b}}(x) &= \int_0^x dy \, f_a(y) H_{\vec{b}}(y) \,, \qquad
    \text{with } f_a = \frac{1}{|a| - \mathop{\mathrm{sgn}}(a) x}
  \end{align}
  In this notation, for example, $f_{-2} = 1/(2+x)$ and $f_2 = 1/(2-a)$.
  Note that for $a > 0$ this differs from the notation in Eq. (80) by an
  overall sign. Obviously, this is a natural generalization of the
  notation of HPLs.
If general real--valued
quantities like mass--ratios or other quantities are present one extends to $a \in 
\mathbb{C}$. Moreover, $\mathfrak{A}_{\rm HPL} \subset \mathfrak{A}_{\rm gHPL}$ holds. In the 
massive OMEs they appeared first in the pure singlet case \cite{Ablinger:2014nga} 
and they contribute also to higher topologies \cite{Ablinger:2014yaa,Ablinger:2015tua}.

The letters which can imply imaginary parts under the transformation $t \rightarrow \pm 
1/x$ are the ones for $a \in \mathbb{R},~|a| \geq 1$. Here, the support of the imaginary 
part is usually not the interval $[0,1]$, as one sees already in the following 
examples.\footnote{Integrals defining $\GA$--functions with singularities in $x 
\in [0,1]$ are dealt with applying Cauchy's valeur principale \cite{CAUCHY}.} 
By defining
\begin{eqnarray}
\gamma_1 = \frac{1}{1 - 2x}
\end{eqnarray}
we consider the following functions
\begin{eqnarray}
\tilde{F}_4(t) &=& 
\GA\left(\frac{1}{2-y}; t\right),
\\ 
\tilde{F}_5(t) &=& \frac{t}{t-1}\Biggl[\HA_{0,0,0,1}\left(t\right) + 
2\GA\left(\gamma_1,1,1,2;t\right)\Biggr], 
\\
\tilde{F}_6(t) &=& 
\frac{t}{t-1}\Biggl[\HA_{0,0,0,1}\left(t\right) +
2\GA\left(1,\gamma_1,1,2;t\right)
+ 2 \GA\left(\gamma_1,1,1,2;t\right)
+ 4 \GA\left(\gamma_1,\gamma_1,1,2;t\right) \Biggr]. 
\end{eqnarray}
Here the index--labels 1 and 2 refer to $1/x$ and $1/(1-x)$, respectively. 
The first terms of their series expansions read
\begin{eqnarray}
\tilde{F}_4(t) &=& \frac{t}{2}+\frac{t^2}{8}+\frac{t^3}{24}+\frac{t^4}{64}+\frac{t^5}{160}
+\frac{t^6}{384}+\frac{t^7}{896}+\frac{t^8}{2048} + O(t^9),
\label{ser:F4}
\\
\tilde{F}_5(t) &=& -t^2-\frac{33 t^3}{16}-\frac{4525 t^4}{1296}-\frac{116929 t^5}{20736}
-\frac{117630361 t^6}{12960000}-\frac{63963307 t^7}{4320000}-\frac{85154778809 t^8}{3457440000}
\nonumber\\ && 
+ O(t^9),
\\
\tilde{F}_6(t) &=& 
-t^2-\frac{41 t^3}{16}-\frac{6685 t^4}{1296}-\frac{199729 t^5}{20736}
-\frac{227246761 t^6}{12960000}-\frac{411349121 t^7}{12960000}-\frac{1792733759681 t^8}{31116960000} 
\nonumber\\ &&
+ O(t^9).
\label{ser:F6}
\end{eqnarray}
In $x$--space one obtains
\begin{eqnarray}
F_4(x) &=& \theta\left(\frac{1}{2} - x \right), 
\\
F_5(x) &=& 
- \frac{1}{1-x}\Biggl\{ \theta (1-x) \Biggl[ \frac{1}{24} \big(
                4 \ln^3(2)
               -2 \ln(2) \pi ^2
                +21 \zeta_3
        \big) 
- \HA_{2,0,0}(x) \Biggr]
\nonumber\\ &&
- \theta(2-x) \frac{1}{24}\big(
        4 \ln^3(2)
        -2 \ln(2) \pi ^2
        +21 \zeta_3
\big)\Biggr\},
\\
F_6(x) &=& - \frac{1}{1-x} \Biggl\{
\theta (1-x) \Biggl[
        \frac{\ln^3(2)}{6}
        +\frac{1}{12} \big(
                -6 \ln^2(2) + \pi ^2\big) \HA_2(x)
        -\frac{1}{8} \zeta_3
\nonumber\\ &&
        + \HA_{2,2,0}(x)
\Biggr]
+\theta (2-x) \Biggl[
        -\frac{\ln^3(2)}{6}
        +\frac{1}{12} \big(
                6 \ln^2(2) - \pi^2\big) \HA_2(x)
        +\frac{1}{8} \zeta_3
\Biggr]
\Biggr\},
\end{eqnarray}
with $\theta$ the Heaviside function. Here regularizations at $x=1$ are necessary.
The transformations used for the functions $F_{4,5,6}$ are not 
part of the package {\tt HarmonicSums}.

If different letters of the kind $1/(x-a),~~a \in ]0,1]$, contribute, there are several cuts contributing to the
$\GA$--functions, which need a closer consideration. The Mellin transform of the functions $F_{5(6)}(x)$
have to be performed using the support $x \in [0,2]$, 
\begin{eqnarray}
\tilde{\Mvec}_a[f(x)](N) = \int_0^a dx x^{N-1} f(x),~~a \in \mathbb{R},
\end{eqnarray}
where the $+$-prescription reads 
\begin{eqnarray}
\tilde{\Mvec}_a^{+,b}[g(x)](N) = \int_0^a dx (x^{N-1} - b^{N-1}) f(x),~~a,b \in \mathbb{R},
\end{eqnarray}
and applies to $b=1$ here. 

The following Mellin transforms are obtained,
\begin{eqnarray}
\Mvec[F_4(x)](N) &=&  \frac{2^{-N}}{N},
\\
\label{eq:test1}
\tilde{\Mvec}_2^{+,1}[F_5(x)](N) &=& - S_{1,3}\left(2, \frac{1}{2}\right)(N-1),
\\
\label{eq:test2}
\tilde{\Mvec}_2^{+,1}[F_6(x)](N) &=& - S_{1,1,2}\left(2,1,\frac{1}{2}\right)(N-1).
\end{eqnarray}
They are in accordance with (\ref{ser:F4}--\ref{ser:F6}). 
The generalized harmonic sums are given by \cite{Ablinger:2013cf}
\begin{eqnarray}
S_{b,\vec{a}}(c,\vec{d})(N) = \sum_{k=1}^N \frac{c^k}{k^b} S_{\vec{a}}(\vec{d})(k),~~b,a_i \in 
\mathbb{N} \backslash \{0\},~~c, d_i \in \mathbb{C} \backslash \{0\}.
\end{eqnarray}

Let us finally note that the generalized harmonic polylogarithms
which occurred in this section can be expressed in terms of harmonic
polylogarithms if we allow for the arguments $x/2$ and $1-x$,
\begin{eqnarray}
\HA_2(x) &=& - \HA_{-1}(1-x) + \ln(2),
\\
\HA_{2,0,0}(x) &=& \frac{1}{2} \Biggl[
        [- \HA_{-1}(1-x) + \ln(2)] \HA_0^2(x)
        -2 \HA_0(x) \HA_{0,1}\left(
                \frac{x}{2}\right)
        +2 \HA_{0,0,1}\left(
                \frac{x}{2}\right)
\Biggr],
\\
\HA_{2,2,0}(x) &=& \frac{1}{2} \ln^2(2) \HA_0(x)
+ \frac{1}{2} \HA_{-1}^2(1-x) \HA_0(x)
+ \left[
        - \ln(2) \HA_0(x)
        + \HA_{0,1}\left(
                \frac{x}{2}\right)
\right] \HA_{-1}(1-x)
\nonumber\\ &&
- \ln(2) \HA_{0,1}\left(
        \frac{x}{2}\right)
+ \HA_{0,1,1}\left(
        \frac{x}{2}\right).
\end{eqnarray}

\subsection{Square root valued alphabets} 
\label{sec:34}

\vspace*{1mm} 
\noindent
Square--root valued alphabets extend those of the previous sections 
by
\begin{eqnarray}
\label{eq:ALSQRT}
\mathfrak{A}_{\rm sqrt} &=& \Biggl\{h_1, h_2, h_3, h_4, h_5, h_6, 
\ldots\Biggr\}
\nonumber\\
\hspace*{-1mm} &=& 
\Biggl\{\frac{1}{x}, \frac{1}{1-x}, \frac{1}{1+x},
\frac{\sqrt{1-x}}{x}, \sqrt{x(1-x)}, \frac{1}{\sqrt{1-x}},
\frac{1}{\sqrt{x} \sqrt{1 \pm x}},
\frac{1}{x \sqrt{1 \pm x}},
\frac{1}{\sqrt{1 \pm x} \sqrt{2 \pm x}},
\nonumber\\ &&
\frac{1}{x \sqrt{1 \pm  x/4}}, ... \Biggr\},
\end{eqnarray}
cf.~\cite{Ablinger:2014bra}. For massive OMEs in the single--mass case theses 
structures appeared first in $A_{gg,Q}$ at three--loop order 
\cite{Ablinger:2014uka,Ablinger:2022wbb}, see also \cite{Ablinger:2015tua}.

Let us consider the following $\GA$--functions,
\begin{eqnarray}
\label{EX7}
\tilde{F}_7(t) &=& \GA\left(4;t\right)
\\
\tilde{F}_8(t) &=& \GA\left(4,2;t \right)
\\
\label{EX9}
\tilde{F}_9(t) &=& 
\GA\left(4,1,2,2;t\right),
\end{eqnarray}
where the index--labels are those of (\ref{eq:ALSQRT}).
Note that G(4;t) has a trailing letter that is singular in the limit
   $t \to 0$. It therefore requires the regularization prescription
   described in Eq. (\ref{eq:REG}). The functions in Eqs.~(\ref{EX7}--\ref{EX9})
have the following series expansions
\begin{eqnarray}
\label{eq:exF7}
\tilde{F}_7(t) &=& -\frac{t}{2}-\frac{t^2}{16}-\frac{t^3}{48}-\frac{5 t^4}{512}-\frac{7 t^5}{1280}
-\frac{7 t^6}{2048}-\frac{33 t^7}{14336}-\frac{429 t^8}{262144} + O(t^9),
\\
\tilde{F}_8(t) &=& t-\frac{t^3}{72}-\frac{t^4}{96}-\frac{71 t^5}{9600}-\frac{31 t^6}{5760}
-\frac{3043 t^7}{752640}-\frac{2689 t^8}{860160} + O(t^9),
\\
\label{eq:exF9}
\tilde{F}_9(t) &=& \frac{t^2}{8}+\frac{t^3}{72}-\frac{t^5}{480}-\frac{881 t^6}{414720}
-\frac{1747 t^7}{967680}-\frac{4561 t^8}{3096576} + O(t^9).
\end{eqnarray}
In $x$--space one obtains
\begin{eqnarray}
F_7(x) &=&
        1
        -\frac{2 (1-x) (1 + 2 x)}{\pi} \sqrt{\frac{1-x}{x}}
        -\frac{8}{\pi} \GA\big(5; x \big),
\\
F_8(x) &=& 
- \frac{1}{\pi}\Biggl[
        4 \frac{(1-x)^{3/2}}{\sqrt{x}}
        +2 (1-x) (1+2 x) \sqrt{\frac{1-x}{x}} [\HA_0 + \HA_1]
        \nonumber \\ &&
        +8 [\GA\big(5,2;x\big) + \GA\big(5,1;x\big)]\Biggr],
\\
F_9(x) &=& 
- \frac{1}{\pi } 
\Biggl\{
        -\biggl[
         16  (1+x)
        +\biggl(
                 8  (1+x)
                +4  (1+x) \HA_1
                \nonumber \\ &&
                +2  (1+2 x) \HA_{0,1}
        \biggr) \HA_0
        +2 (1+x) \HA_0^2
        +\frac{1}{3} (1+2 x) \HA_0^3
        +8 (1+x) \HA_1
        \nonumber \\ &&
        +2 (1+x) \HA_1^2
        -2 (1+2 x) \HA_{0,0,1}
        +2 (1+2 x) \HA_{0,1,1}
        \biggr] (1-x) \sqrt{\frac{1-x}{x}}
        \nonumber \\ &&
        +\biggl(
                 12 (1-x) (1+x) \sqrt{\frac{1-x}{x}}
                +6 (1-x) (1+2 x) \sqrt{\frac{1-x}{x}} \HA_0
                +36 {\rm G}\big(5; x\big)
\nonumber\\ &&                 +24 {\rm G}\big(5,1; x\big)
        \biggr) \zeta_2
        +\biggl(
                 2 (1-x) (1+2 x) \sqrt{\frac{1-x}{x}}
                +8 {\rm G}\big(5; x\big)
        \biggr) \zeta_3
        -32 {\rm G}\big(5; x\big)
        \nonumber \\ &&
        -16 {\rm G}\big(5,2; x\big)
        -16 {\rm G}\big(5,1; x\big)
        -12 {\rm G}\big(5,2,2; x\big)
        -12 {\rm G}\big(5,2,1; x\big)
        \nonumber \\ &&
        -12 {\rm G}\big(5,1,2; x\big)
        -12 {\rm G}\big(5,1,1; x\big)
        -8 {\rm G}\big(5,1,2,1; x\big)
        -8 {\rm G}\big(5,1,2,2; x\big)
        \nonumber \\ &&
        -8 {\rm G}\big(5,1,1,1; x\big)
        -8 {\rm G}\big(5,1,1,2; x\big)
\Biggr\}.
\end{eqnarray}
The Mellin transforms of the above examples for general values of $N$ will also 
contain cyclotomic harmonic sums \cite{Ablinger:2011te}  and central 
binomial terms
\cite{Ablinger:2014bra}. The inversion to $x$--space has been performed by solving differential 
equations. The corresponding Mellin transforms read
\begin{eqnarray}
\Mvec[F_7(x)](N) &=& -\frac{2^{1-2 N}}{N^2} \binom{2 N - 2}{N-1},
\\
\Mvec[F_8(x)](N) &=& 
-\frac{\binom{2 N}{N}}{2^{2 N-1} N (2 N - 1)}
        S_{\{2,-3,1\}}({N})
\\
\Mvec[F_9(x)](N) &=& 
 \frac{\binom{2 N}{N}}{2^{2 N}} \Biggl[
           \frac{16 \big(-1-4 N-32 N^2+16 N^3+16 N^4\big)}{(-1+2 N)^4 (1+2 N)^3}
           +\frac{4 S_{\{2,1,1\}}^3({N})}{3 N (-1+2 N)}
  \nonumber \\ &&
           +\biggl(
             -\frac{16 \big(-1-8 N+4 N^2\big)}{(-1+2 N)^3 (1+2 N)^2}
             -\frac{4 S_{\{2,1,2\}}({N})}{N (-1+2 N)}
           \biggr) S_{\{2,1,1\}}({N})
  \nonumber \\ && 
           -\frac{16 (2+N) (-1+8 N) S_{\{2,1,1\}}^2({N})}{15 N (-1+2 N)^2 (1+2 N)}
           -\frac{4 S_{\{1,0,1\},\{2,1,1\},\{2,1,1\}}({N})}{N (-1+2 N)}
  \nonumber\\ && 
           -\frac{16 (-2+N) (1+8 N) S_{\{2,1,2\}}({N})}{15 N (-1+2 N)^2 (1+2 N)}
           +\frac{64 S_{\{2,1,1\},\{2,1,1\}}({N})}{15 N (-1+2 N)}
  \nonumber \\ &&
           +\frac{4 S_{\{1,0,1\},\{2,1,2\}}({N})}{N (-1+2 N)}
           +\frac{8 S_{\{2,1,3\}}({N})}{3 N (-1+2 N)}
         \Biggr],
\end{eqnarray}
and agree with the coefficients of the expansions (\ref{eq:exF7}--\ref{eq:exF9}). Here the cyclotomic sums are
\begin{eqnarray}
S_{\{a_1,a_2,a_3\},\{
\vec{b}_1,
\vec{b}_3,
\vec{b}_3\}}(N) = \sum_{k=1}^N \frac{1}{(a_1 k + a_2)^{a_3}} S_{\{
\vec{b}_1,
\vec{b}_3,
\vec{b}_3\}}(k).
\end{eqnarray}
Note that for root--valued iterated integrals letters containing factors
\begin{eqnarray}
(1 \pm t)^\alpha,~~~\alpha \in \mathbb{R},
\end{eqnarray}
may imply the occurrence of an imaginary part after transforming $t \rightarrow \pm 1/x$, which 
generalizes the 
case of the letter $1/(1 \pm t)$ in the previous classes of functions. Furthermore, for more general root 
valued letters, cf. \cite{Ablinger:2014bra}, also other cuts need to be considered.

In very simple cases the integrals defining $\GA$--functions lead to known functions, cf.~\cite{Ablinger:2022wbb}
for a series of examples. In particular at higher depth also special constants contribute, which can be calculated 
using methods for infinite binomial sums \cite{Ablinger:2014bra,Davydychev:2003mv,Weinzierl:2004bn}.
\section{Iterated non--iterative integrals} 
\label{sec:4}

\vspace*{1mm} 
\noindent
Beyond the purely iterated  integrals, there are also integrals, which cannot be written in this way.
Instead of iterated integrals over alphabets of rational or irrational functions, the 
respective letters are given by higher transcendental functions which are themselves 
defined by at least one definite integral. 
Its $x$--dependence comes from an argument of the integrand and cannot 
be transformed to only the boundary of the integral.
The simplest cases of this kind found 
in physics applications seem to be so-called $_2F_1$--solutions. In the case we consider 
in the following it turns out that
the hierarchy of master integrals is such 
that the $_2F_1$--solutions occur only in the seeds 
and the other master integrals are given by first--order iterations
over them.
For this reason we called these integrals iterated non--iterative integrals \cite{INII}. 
This class also covers a wide range of concrete
cases which occur in Feynman diagram calculations such as Abel integrals \cite{Neumann}, $K3$ surfaces
\cite{Brown}, and Calabi--Yau motives \cite{CY}, see also Ref.~\cite{ELLREV}.
We will first consider the basic $_2F_1$--solutions emerging in the massive OME
$A_{Qg}$, find solutions of the corresponding master integrals in a Laurent expansion  in $\ep$, 
and derive the $x$--space representation for these non--iterative master integrals in Section~\ref{sec:41}. 
In  Section~\ref{sec:42} we describe the principal method to iteratively determine 
higher master integrals, which depend on $_2F_1$--solutions in their inhomogeneous 
part.
\subsection{\boldmath $_2F_1$ solutions}
\label{sec:41}

\vspace*{1mm} 
\noindent
We consider the six master integrals leading to $_2F_1$--solutions and contributing to the massive OME 
$A_{Qg}^{(3)}$, cf.~\cite{Bierenbaum:2009mv}. 
They are given by
  \begin{align}
    \mathsf{F}_1(t)
      &= \frac{1}{(2\pi)^{3D}} \iiint
         \frac{d^D k_1 \, d^D k_2 \, d^D k_3}{D_1 D_4 D_6 D_7 D_{10}}
    \,, \\
    \mathsf{F}_2(t)
      &= \frac{1}{(2\pi)^{3D}} \iiint
         \frac{d^D k_1 \, d^D k_2 \, d^D k_3}{D_1^2 D_4 D_6 D_7 D_{10}}
    \,, \\
    \mathsf{F}_3(t)
      &= \frac{1}{(2\pi)^{3D}} \iiint
         \frac{d^D k_1 \, d^D k_2 \, d^D k_3}{D_1^3 D_4 D_6 D_7 D_{10}}
    \,, \\
    \mathsf{F}_4(t)
      &= \frac{1}{(2\pi)^{3D}} \iiint
         \frac{d^D k_1 \, d^D k_2 \, d^D k_3}{D_2 D_3 D_6 D_7 D_{10}}
    \,, \\
    \mathsf{F}_5(t)
      &= \frac{1}{(2\pi)^{3D}} \iiint
         \frac{d^D k_1 \, d^D k_2 \, d^D k_3}{D_2^2 D_3 D_6 D_7 D_{10}}
    \,, \\
    \mathsf{F}_6(t)
      &= \frac{1}{(2\pi)^{3D}} \iiint
         \frac{d^D k_1 \, d^D k_2 \, d^D k_3}{D_2^3 D_3 D_6 D_7 D_{10}}
    \,,
  \end{align}
  and the propagators read
\begin{align}
    D_1 &= k_1^2-m^2 \,, &
    D_2 &= (k_1-p)^2-m^2 \,, \\
    D_3 &= k_2^2-m^2 \,, &
    D_4 &= (k_2-p)^2-m^2 \,, \\
    D_6 &= (k_1-k_3)^2-m^2 \,, &
    D_7 &= (k_2-k_3)^2-m^2 \,, \\
    D_{10} &= 1 - t (\Delta.k_1) \,.
  \end{align}
with $m$ a heavy quark mass.
  The three integrals $\mathsf{F}_{4,5,6}(t)$ are related to
  $\mathsf{F}_{1,2,3}(t)$, respectively, by conjugation and, therefore,
  do not need to be calculated by solving the associated differential
  equations.
The remaining system of three first--order differential equations can be decoupled by {\tt 
OreSys} \cite{DECOUP,ORESYS} into one differential equation of order {\sf o = 3} and two 
differential relations for the  other
functions ${\sf F}_k(t),~~k \in \{1,2,3\}$. 
The original system of differential equations has the following coefficient matrix
\renewcommand{\arraystretch}{1.3}
\begin{eqnarray}
M_1(t,\ep) =
\left[
\begin{array}{rrr}
-\frac{1}{t} & - \frac{1}{1-t} & 0 \\
           0 & - \frac{1}{t(1-t)} & -\frac{2}{1-t} \\
           0 &   \frac{2}{t(8+t)} &  \frac{1}{8+t} 
\end{array}
\right]
+ \ep \left[
\begin{array}{rrr}
-\frac{1}{2 t} & 0 & 0 \\
           0 & - \frac{1}{2 t} & 0 \\
           -\frac{(1-t)}{2 t (8+t)}\left[1 + \frac{7 \ep}{4} +\frac{3 \ep^2}{8}\right] & 
\frac{2(13 - 4 t) - \ep (7 + 11 t)}{8 t (8+t)}
&
\frac{16 + 5 t}{2 t (8+t)}
\end{array}
\right]
\nonumber\\
\end{eqnarray}
\renewcommand{\arraystretch}{1.0}
and it is given by
\renewcommand{\arraystretch}{1.3}
\begin{eqnarray}
\frac{d}{dt} 
\left[\begin{array}{c} \F_1(t, \ep) \\ \F_2(t, \ep) \\ \F_3(t, \ep)\end{array}\right]
= M_1(t,\ep)
\left[\begin{array}{c} \F_1(t, \ep) \\ \F_2(t, \ep) \\ \F_3(t, \ep)\end{array}\right]
+ \left[\begin{array}{c} R_1(t,\ep) \\ R_2(t,\ep) \\ R_3(t,\ep)\end{array}\right] + O(\ep),
\label{eq:DIEQS}
\end{eqnarray}
\renewcommand{\arraystretch}{1.0}
\noindent
where the inhomogeneities are
\begin{eqnarray}
R_1(t,\ep) &=& \frac{1}{t(1-t)\ep^3}\left[
16 - \frac{68}{3}\ep  +  \left(\frac{59}{3} + 6 \zeta_2\right) \ep^2 + \left(
-\frac{65}{12} - \frac{17}{2} \zeta_2 + 2 \zeta_3\right) \ep^3\right] + O(\ep),
\nonumber\\
\\
R_2(t,\ep) &=& \frac{1}{t(1-t) \ep^3} \left[
8
-\frac{16}{3} \ep
+\left(
        \frac{4}{3}
        +3 \zeta_2
\right) \ep^2
+ \left(
        \frac{14}{3}
        -2 \zeta_2
        +\zeta_3
\right) \ep^3
\right]
+ O(\ep),
\\
R_3(t,\ep) &=& \frac{1}{12 t(8+t) \ep^3}\left[-192
+8 \ep
-8 \big(
        4
        +9 \zeta_2
\big) \ep^2
+ \big(
        68
        +3 \zeta_2
        -24 \zeta_3
\big) \ep^3 \right]
+ O(\ep).
\end{eqnarray}
The functions $\F_i(t, \ep)$ are expanded into a Laurent series in $\ep$,
\begin{eqnarray}
\F_i(t, \ep) = \sum_{k = -3}^{\infty} \F_{i,k}(t) \ep^k.
\end{eqnarray}
We first solve the homogeneous system after the decoupling for one of the functions $\F_i$ is performed. Then
the differential equations will be solved by using the method presented in Ref.~\cite{Ablinger:2018zwz}
looping up in the dimensional parameter $\ep$. Here also decoupling is used, cf.~Ref.~\cite{ORESYS}.

Concerning the simplicity of the solution structure, it is important for which of the functions one decouples 
first. If one chooses $\F_1$, see Appendix~\ref{sec:B}, a more complicated structure is obtained than starting with 
$\F_3$. The former case is structurally closer to the solution found in Ref.~\cite{Ablinger:2017bjx}.
In Appendix~\ref{sec:B} we show the lengthy expression of the solution $\F_1(t)$ up to $O(\ep^{-1})$, which 
is given by $\GA$--functions containing $_2F_1$--letters in a spurious manner. Actually, a much more compact 
solution,
free of $_2F_1$--letters, is obtained, as will be shown in Eq.~(\ref{eq:F1A}). The reason for this is, that 
the original $3 \times 3$ system has been transformed into a third--order differential equation without factorizing 
into a first--order and a second--order system first and solving first the first--order equation. 

One is generally advised to solve first the differential equations of the first--order sub--systems.\footnote{In 
Mellin space the package {\tt Sigma}
\cite{SIG1,SIG2} always factorizes first all first--order factors. This is generally not the case
for decoupling algorithms \cite{DECOUP} implemented in {\tt OreSys} \cite{ORESYS}. However,
one can investigate differential equation decoupling using e.g. the algorithm \cite{DEfact}
available in {\tt Maple}.} 
If we decouple for the solution of $\F_3(t)$   using {\tt OreSys} first 
we obtain the homogeneous differential equation
\begin{eqnarray}
\F'_1(t) +  \frac{1}{t} \F_1(t) = 0. 
\end{eqnarray}
The other solutions appear only in the inhomogeneity. The particular solution of the homogeneous equation is  
\begin{eqnarray}
\tilde{g}_0(t) = \frac{1}{t}.
\end{eqnarray}
Further, the homogeneous differential equation of $\F_3(t)$ is now given by 
\begin{eqnarray}
\label{eq:F1new}
\F_3''(t) + \frac{(2 - t)}{(1 - t) t} \F_3'(t) + \frac{2+t}{(1-t)t(8+t)} \F_3(t) = 0, 
\end{eqnarray}
while the solution $\F_2(t)$ is a function of $\F_3(t)$ and its derivatives. In this way, the $3 \times 3$ system
decouples into a first--order and a second--order system. In general, one is advised to find all first--order 
solutions 
through decoupling of the complete system first.

The Heun equation \cite{HEUN} (\ref{eq:F1new}) has singularities at $t_0 \in \{ -8, 0, 1, \infty \}$. They will 
transform into 
$x_0 \in \{-1/8, 0, 1, \infty \}$ and one therefore expects that the series around $x=0$ has a convergence 
radius $r < 1/8$, which has consequences for the final numerical representation. Eq.~(\ref{eq:F1new}) has 
the advantage that there are no singularities in $x \in 
]0,1[$, unlike the case of the elliptic solutions in \cite{Ablinger:2017bjx}, Eqs.~(3.18, 3.19), or
Eqs.~(\ref{eq:HEUN3a}, \ref{eq:HEUN3b}), providing an easier way to perform the analytic continuation.

The pair of particular solutions of the homogeneous equation  Eq.~(\ref{eq:F1new}) is given by
\begin{eqnarray}
\label{eq:gtil1}
\tilde{g}_1(t) &=&  
\frac{2}{(1-t)^{2/3} (8+t)^{1/3}} 
\pFq{2}{1}{\frac{1}{3}, \frac{4}{3}}{2}{-\frac{27 t}{(1-t)^2 (8+t)}},
\\
\label{eq:gtil2}
\tilde{g}_2(t) &=& \frac{9 \sqrt{3} \Gamma^2(1/3)}{8\pi}  \frac{1}{(1-t)^{2/3} (8+t)^{1/3}} 
\pFq{2}{1}{\frac{1}{3}, \frac{4}{3}}{\frac{2}{3}}{1+\frac{27 t}{(1-t)^2 (8+t)}},
\end{eqnarray}
with the Wronskian
\begin{eqnarray}
W(t) = \frac{1-t}{t^2}.
\end{eqnarray}
The normalization of $\tilde{g}_2(t)$ has been chosen in such a way that the Wronskian is free
of transcendental constants.
Note that the parameters of the $_2F_1$--functions are not the same as in 
Eqs.~(\ref{eq:HEUN3a}, \ref{eq:HEUN3b}). In the solutions also the functions $\tilde{g}'_{1(2)}(t)$ are 
contributing, while higher derivatives are expressed using Eq.~(\ref{eq:F1new}).
The functions $\tilde{g}_{1(2)}(t)$ are discontinuous at $t=1$,
\begin{alignat}{3}
\lim_{t \rightarrow 1^-} {\sf Re}[\tilde{g}_{1}(t)] &= \frac{3 \sqrt{3}}{2 \pi}, 
&~~~\lim_{t \rightarrow 1^-} {\sf Re}[\tilde{g}_{2}(t)] 
&= \frac{9}{8},
\\
\lim_{t \rightarrow 1^+} {\sf Re}[\tilde{g}_{1}(t)] &= - \frac{3 \sqrt{3}}{4 \pi}, 
&~~~\lim_{t \rightarrow 1^+} {\sf Re}[\tilde{g}_{2}(t)]
&= - \frac{9}{4},
\\
\lim_{t \rightarrow 1^-} {\sf Im}[\tilde{g}_{1}(t)] &= 0,
&~~~\lim_{t \rightarrow 1^-} {\sf Im}[\tilde{g}_{2}(t)] &= 
- \frac{9 \sqrt{3}}{8},
\\
\lim_{t \rightarrow 1^+} {\sf Im}[\tilde{g}_{1}(t)] &= - \frac{9}{4 \pi}, 
&~~~\lim_{t \rightarrow 1^+} {\sf Im}[\tilde{g}_{2}(t)] &= 0.
\end{alignat}
This requires to consider the cases $t < 1$ and $t > 1$ separately.

The solutions $\F_i(t)$ of the $3 \times 3$ system up to $O(\ep^0)$ can be expressed as 
iterated integrals over the alphabet
\begin{eqnarray}
\label{eq:ALPH2}
\mathfrak{A}_2 &=& 
\Biggl\{
\frac{1}{t},
\frac{1}{1-t},
\frac{1}{8+t},
\tilde{g}_1,                  
\tilde{g}_2,                  
\frac{\tilde{g}_1}{t},        
\frac{\tilde{g}_1}{1-t},      
\frac{\tilde{g}_1}{8+t},      
\frac{\tilde{g}_1'}{t},       
\frac{\tilde{g}_1'}{1-t},     
\frac{\tilde{g}_1'}{8+t},     
\frac{\tilde{g}_2}{t},        
\frac{\tilde{g}_2}{1-t},      
\frac{\tilde{g}_2}{8+t},      
\frac{\tilde{g}_2'}{t},       
\frac{\tilde{g}_2'}{1-t},     
\nonumber\\ &&
\frac{\tilde{g}_2'}{8+t},     
t \tilde{g}_1,                
t \tilde{g}_2                 
\Biggr\}
\end{eqnarray}
of length 19. Later we will refer to $\GA$--functions also for $x \in [0,1]$. The corresponding alphabet 
is obtained by setting $t \rightarrow 1/x$ and partial fractioning.
For technical reasons additional regularization may 
become necessary later
because of the small--$t$ behaviour of these letters.

In the $\GA$--functions below the respective letter
is denoted by its position in $\mathfrak{A}_2$. One might express $\tilde{g}'_2$ by  
\begin{eqnarray}
\tilde{g}_2' = \frac{1}{\tilde{g}_1} \left[\gx_2 \gx_1'  + \frac{1}{t^2} - \frac{1}{t}\right],
\end{eqnarray}
which we will not apply, however, since $\tilde{g}_1$ would appear in the denominator, which is technically 
more difficult in some representations.

The system relates to all solutions ${\sf F}_i(t)$ through the inhomogeneities. At higher order in $\ep$ all 
solutions obtain $\GA$--functions containing $_2F_1$--dependent letters.
We first compute the functions ${\sf F}_i(t)$ in the region $t \in [0,1^-]$. The initial conditions are set at 
$t=0$.
From these solutions one can calculate the associated analytic expansion around $x = 1$.

To $O(\ep^0)$ the solutions read
\begin{eqnarray}
\label{eq:F1A}
\F_1(t) &=& 
  \frac{8}{\ep^3} \left[1 + \frac{1}{t} \HA_1(t)\right]
- \frac{1}{\ep^2} \Biggl[
        \frac{1}{6} (106+t)
        +\frac{(9+2 t)}{t} \HA_1(t)
        +\frac{4}{t} \HA_{0,1}(t)
\Biggr]
\nonumber\\ &&
+\frac{1}{\ep} \Biggl\{
        \frac{1}{12} (271+9 t)
        +\Biggl[
                \frac{71+32 t+2 t^2}{12 t}
                +\frac{3 \zeta_2}{t}
        \Biggr] \HA_1(t)
        +\frac{(9+2 t)}{2 t} \HA_{0,1}(t)
        +\frac{2}{t} \HA_{0,0,1}(t)
\nonumber\\ &&       
 +3 \zeta_2 \Biggr\}
+ \frac{1}{t} \Biggl\{
        \frac{6696-22680 t-16278 t^2-255 t^3-62 t^4}{864 t}
        +\big(
                9+9 t+t^2\big) \gx_1(t) \Biggl[
                \frac{31 \ln(2)}{16}
\nonumber\\ &&              
  +\frac{1}{144} \big(
                        265+31 \pi  (-3 i + \sqrt{3})
\big)
                +\frac{3}{8} \ln(2) \zeta_2
                +\frac{1}{24} \big(
                        10+\pi  (-3 i + \sqrt{3})
\big) \zeta_2
                -\frac{7}{4} \zeta_3
        \Biggr]
\nonumber\\ && 
        +\GA(18,t) \Biggl[
                -\frac{93 \ln(2)}{16}
                +\frac{1}{48} \big(
                        -265-31 \pi  (-3 i + \sqrt{3})
\big)
                +\Biggl(
                        -\frac{9 \ln(2)}{8}
\nonumber\\ &&               
         +\frac{1}{8} \big(
                                -10-\pi  \big(
                                        -3 i
                                        +\sqrt{3}
                                \big)\big)
                \Biggr) \zeta_2
                +\frac{21}{4} \zeta_3
        \Biggr]
        +\GA(16;t) \Biggl[
                \frac{31}{4}
                +\frac{3}{2} \zeta_2
                +\big(
                        9+9 t+t^2
                \big)
\nonumber\\ && 
\Biggl(\frac{31}{36}
                        +\frac{\zeta_2}{6}
                \Biggr) \gx_1(t)
        \Biggr]
        +\GA(13;t) \Biggl[
                -\frac{31}{36}
                -\frac{1}{6} \zeta_2
                +\big(
                        9+9 t+t^2
                \big)
\Biggl(\frac{655}{648}
                        +\frac{25 \zeta_2}{108}
                \Biggr) \gx_1(t)
        \Biggr]
\nonumber\\ && 
        +\GA(4;t) \Biggl[
                -\frac{155 \ln(2)}{8}
                -\frac{5}{72} \big(
                        265+31 \pi  (-3 i + \sqrt{3})
\big)
                +\Biggl(
                        -\frac{15 \ln(2)
                        }{4}
\nonumber\\ &&                
         -
                        \frac{5}{12} \big(
                                10+\pi  \big(
                                        -3 i
                                        +\sqrt{3}
                                \big)\big)
                \Biggr) \zeta_2
                +\frac{35}{2} \zeta_3
                -\frac{7}{24} \big(
                        9+9 t+t^2\big) \gx_2(t)
        \Biggr]
        +\GA(7;t) \Biggl[
                \frac{31 \ln(2)}{16}
\nonumber\\ &&                
 +\frac{1}{144} \big(
                        265+31 \pi  (-3 i + \sqrt{3})
\big)
                +\Biggl(
                        \frac{3 \ln(2)}{8}
                        +\frac{1}{24} \big(
                                10+\pi  \big(
                                        -3 i
                                        +\sqrt{3}
                                \big)\big)
                \Biggr) \zeta_2
                -\frac{7}{4} \zeta_3
\nonumber\\ &&                
 -\big(
                        9+9 t+t^2
                \big)
\Biggl(\frac{655}{648}
                        +\frac{25 \zeta_2}{108}
                \Biggr) \gx_2(t)
        \Biggr]
        +\GA(10;t) \Biggl[
                -\frac{279 \ln(2)}{16}
                +\frac{1}{16} \big(
                        -265
\nonumber\\ && 
-31 \pi  (-3 i + \sqrt{3})
\big)
                +\Biggl(
                        -\frac{27 \ln(2)}{8}
                        -\frac{3}{8} \big(
                                10+\pi  \big(
                                        -3 i
                                        +\sqrt{3}
                                \big)\big)
                \Biggr) \zeta_2
                +\frac{63}{4} \zeta_3
\nonumber\\ &&                
 -\frac{31}{36} \big(
                        9+9 t+t^2\big) \gx_2(t)
                -\frac{1}{6} \big(
                        9+9 t+t^2\big) \zeta_2 \gx_2(t)
        \Biggr]
        -\Biggl(
                \frac{31}{4}
                +\frac{3 \zeta_2}{2}
        \Biggr) \HA_0(t)
\nonumber\\ &&    
     -\Biggl(
                \frac{1}{144} \big(
                        809+564 t+75 t^2+4 t^3\big)
                +\frac{1}{4} (23+3 t) \zeta_2
                -\zeta_3
        \Biggr) \HA_1(t)
        -\Biggl(
                \frac{1}{24}
                 \big(
                        71
\nonumber\\ && 
+24 t-3 t^2\big)
                +
                \frac{3 \zeta_2}{2}
        \Biggr) \HA_{0,1}(t)
        -\frac{1}{4} (9+2 t) \HA_{0,0,1}(t)
        -\HA_{0,0,0,1}(t)
        +\frac{1}{4} (63+4 t) \zeta_3
\nonumber\\ && 
        +\frac{\big(
                12-45 t-46 t^2+3 t^3\big) \zeta_2}{8 t}
        -\Biggl(\frac{31}{36} + \frac{\zeta_2}{6} \Biggr)\big(
                9+9 t+t^2\big) \gx_2(t)
\nonumber\\ && 
        +\Biggl(
                \frac{155}{18}
                +\frac{7}{24} \big(
                        9+9 t+t^2\big) \gx_1(t)
                +\frac{5 \zeta_2}{3}
        \Biggr) \GA(5;t)
        +\big(
                9+9 t+t^2
        \big)
\Biggl(\frac{259}{81}
                +\frac{14 \zeta_2}{27}
        \Biggr) \gx_2(t) 
\nonumber\\ && 
\times
\GA(8;t)
        -\big(
                9+9 t+t^2
        \big)
\Biggl(\frac{259}{81}
                +\frac{14 \zeta_2}{27}
        \Biggr) \gx_1(t) \GA(14;t)
        +\Biggl(
                \frac{31}{12}
                +\frac{\zeta_2}{2}
        \Biggr) \GA(19;t)
\nonumber\\ &&   
     -\frac{1}{6} \big(
                9+9 t+t^2\big) \gx_2(t) \GA(4,2;t)
        -\frac{35}{12} \GA(4,5;t)
        -\Biggl(
                \frac{3275}{324}
                +\frac{125 \zeta_2}{54}
        \Biggr) \GA(4,13;t)
\nonumber\\ &&  
       +\Biggl(
                \frac{2590}{81}
                +\frac{140 \zeta_2}{27}
        \Biggr) \GA(4,14;t)
        -\Biggl(
                \frac{155}{18}
                +\frac{5 \zeta_2}{3}
        \Biggr) \GA(4,16;t)
        +\frac{1}{6} \big(
                9+9 t+t^2\big) \gx_1(t) 
\nonumber\\ && 
\times \GA(5,2;t)
        +\frac{35}{12} \GA(5,4;t)
        +\Biggl(
                \frac{3275}{324}
                +\frac{125 \zeta_2}{54}
        \Biggr) \GA(5,7;t)
        -\Biggl(
                \frac{2590}{81}
                +\frac{140 \zeta_2}{27}
        \Biggr) \GA(5,8;t)
\nonumber\\ &&     
    +\Biggl(
                \frac{155}{18}
                +\frac{5 \zeta_2}{3}
        \Biggr) \GA(5,10;t)
        +\frac{1}{24} \big(
                9+9 t+t^2\big) \gx_2(t) \GA(6,2;t)
        +\frac{7}{24} \GA(7,5;t)
\nonumber\\ &&   
     +\Biggl(
                \frac{655}{648}
                +\frac{25 \zeta_2}{108}
        \Biggr) \GA(7,13;t)
        -\Biggl(
                \frac{259}{81}
                +\frac{14 \zeta_2}{27}
        \Biggr) \GA(7,14;t)
        +\Biggl(
                \frac{31}{36}
                +\frac{\zeta_2}{6}
        \Biggr) \GA(7,16;t)
\nonumber\\ &&        
 +\frac{7}{8} \big(
                9+9 t+t^2\big) \gx_2(t) \GA(8,2;t)
        -\frac{21}{8} \GA(10,5;t)
        -\Biggl(
                \frac{655}{72}
                +\frac{25 \zeta_2}{12}
        \Biggr) \GA(10,13;t)
\nonumber\\ &&    
     +\Biggl(
                \frac{259}{9}
                +\frac{14 \zeta_2}{3}
        \Biggr) \GA(10,14;t)
        -\Biggl(
                \frac{31}{4}
                +\frac{3 \zeta_2}{2}
        \Biggr) \GA(10,16;t)
        -\frac{1}{24} \big(
                9+9 t+t^2\big) \gx_1(t) 
\nonumber\\ && 
\times \GA(12,2;t)
        -\frac{7}{24} \GA(13,4;t)
        -\Biggl(
                \frac{655}{648}
                +\frac{25 \zeta_2}{108}
        \Biggr) \GA(13,7;t)
        +\Biggl(
                \frac{259}{81}
                +\frac{14 \zeta_2}{27}
        \Biggr) \GA(13,8;t)
\nonumber\\ && 
        -\Biggl(
                \frac{31}{36}
                +\frac{\zeta_2}{6}
        \Biggr) \GA(13,10;t)
        -\frac{7}{8} \big(
                9+9 t+t^2\big) \gx_1(t) \GA(14,2;t)
        +\frac{21}{8} \GA(16,4;t)
\nonumber\\ &&     
    +\Biggl(
                \frac{655}{72}
                +\frac{25 \zeta_2}{12}
        \Biggr) \GA(16,7;t)
        -\Biggl(
                \frac{259}{9}
                +\frac{14 \zeta_2}{3}
        \Biggr) \GA(16,8;t)
        +\Biggl(
                \frac{31}{4}
                +\frac{3 \zeta_2}{2}
        \Biggr) \GA(16,10;t)
\nonumber\\ &&    
     -\frac{7}{8} \GA(18,5;t)
        -\Biggl(
                \frac{655}{216}
                +\frac{25 \zeta_2}{36}
        \Biggr) \GA(18,13;t)
        +\Biggl(
                \frac{259}{27}
                +\frac{14 \zeta_2}{9}
        \Biggr) \GA(18,14;t)
\nonumber\\ && 
        -\Biggl(
                \frac{31}{12}
                +\frac{\zeta_2}{2}
        \Biggr) \GA(18,16;t)
        +\frac{7}{8} \GA(19,4;t)
        +\Biggl(
                \frac{655}{216}
                +\frac{25 \zeta_2}{36}
        \Biggr) \GA(19,7;t)
        -\Biggl(
                \frac{259}{27}
\nonumber\\ &&                
 +\frac{14 \zeta_2}{9}
        \Biggr) \GA(19,8;t)
        +\Biggl(
                \frac{31}{12}
                +\frac{\zeta_2}{2}
        \Biggr) \GA(19,10;t)
        +\frac{5}{3} [\GA(5,4,2;t)
        - \GA(4,5,2;t)]
\nonumber\\ && 
        +\frac{5}{12} [\GA(4,12,2;t)
        - \GA(5,6,2;t)]
        +\frac{35}{4} [\GA(4,14,2;t)
        - \GA(5,8,2;t)]
\nonumber\\ && 
       +\frac{1}{6} [\GA(7,5,2;t)
        - \GA(13,4,2;t)]
        +\frac{1}{24} [\GA(13,6,2;t)
        - \GA(7,12,2;t)]
\nonumber\\ &&  
      +\frac{1}{4} \big(
                9+9 t+t^2\big) [\gx_2(t) \GA(8,1,2;t)
        - \gx_1(t) \GA(14,1,2;t)]
        +\frac{3}{2} [\GA(16,4,2;t)
        - \GA(10,5,2;t)]
\nonumber\\ &&   
     +\frac{7}{8} [\GA(13,8,2;t)
        - \GA(7,14,2;t)]
        +\frac{3}{8} [\GA(10,12,2;t)
        - \GA(16,6,2;t)]
\nonumber\\ &&  
      +\frac{63}{8} [\GA(10,14,2;t)
        - \GA(16,8,2;t)]
        +\frac{1}{2} [\GA(19,4,2;t)
        - \GA(18,5,2;t)]
\nonumber\\ &&
        +\frac{1}{8} [\GA(18,12,2;t)
        - \GA(19,6,2;t)]
        +\frac{21}{8} [\GA(18,14,2;t)
        - \GA(19,8,2;t)]
\nonumber\\ && 
       +\frac{5}{2} [\GA(4,14,1,2;t)
        - \GA(5,8,1,2;t)]
        +\frac{1}{4} [\GA(13,8,1,2;t)
        - \GA(7,14,1,2;t)]
\nonumber\\ &&  
      +\frac{9}{4} [\GA(10,14,1,2;t)
        - \GA(16,8,1,2;t)]
        +\frac{3}{4} [\GA(18,14,1,2;t)
        - \GA(19,8,1,2;t)]
\Biggr\}
\nonumber\\ &&
+  O(\ep),
\\
\F_2(t) &=& \frac{8}{\ep^3} + \frac{1}{\ep^2}\Biggl[-\frac{1}{3}(34 + t) + \frac{2 (1 - t)}{t} 
\HA_1(t) 
\Biggr]
+ \frac{1}{\ep} \Biggl[
\frac{116 + 15 t}{12} 
+ 3 \zeta_2 
- \frac{(1 - t) (8 + t)}{3t}  \HA_1(t)
\nonumber\\ &&
- \frac{1 - t}{t} \HA_{0,1}(t) 
\Biggr] 
+ \frac{992-368 t+75 t^2-27 t^3}{144 t}
+(1-t) \Biggl(
        \frac{\big(
                43+10 t+t^2\big)}{12 t} \HA_1(t)
        +\frac{(4-t) }{4 t} 
\nonumber\\ && 
\times \HA_{0,1}(t)
        +\frac{3  \zeta_2}{4 t} \HA_1(t)
\Biggr)
+t \Biggl[
                \frac{31 \ln(2)}{16}
                +\frac{1}{144} \big(
                        265
\nonumber\\ &&
+ 31 \pi  \big(
                                - 3 i
                                + \sqrt{3}
                        \big)\big) 
 +\Biggl(
                        \frac{3 \ln(2)}{8}
                        +\frac{1}{24} \big(
                                10+\pi  \big(
                                        -3 i
                                        +\sqrt{3}
                                \big) \Biggr)
                \Biggr) \zeta_2
                -\frac{7}{4} \zeta_3
                +\frac{7}{24} \GA(5;t)
\nonumber\\ &&
                +\Biggl(
                        \frac{655}{648}
                        +\frac{25 \zeta_2}{108}
                \Biggr) \GA(13;t)
                -\Biggl(
                        \frac{259}{81}
                        +\frac{14 \zeta_2}{27}
                \Biggr) \GA(14;t)
                +\Biggl(
                        \frac{31}{36}
                        +\frac{\zeta_2}{6}
                \Biggr) \GA(16;t)
\nonumber\\ && 
                +\frac{1}{6} \GA(5,2;t)
                -\frac{1}{24} \GA(12,2;t)
                -\frac{7}{8} \GA(14,2;t)
                -\frac{1}{4} \GA(14,1,2;t)
        \Biggr]
[-\gx_1(t)
\nonumber\\ &&                
 +(8+t) \gx_1'(t)
        ]
+t \Biggl[
                -
                \frac{31}{36}
                -\frac{1}{6} \zeta_2
                -\frac{7}{24} \GA(4;t)
                -\Biggl(
                        \frac{655}{648}
                        +\frac{25 \zeta_2}{108}
                \Biggr) \GA(7;t)
                +\Biggl(
                        \frac{259}{81}
\nonumber\\ &&                       
  +\frac{14 \zeta_2}{27}
                \Biggr) \GA(8;t)
                -\Biggl(
                        \frac{31}{36}
                        +\frac{\zeta_2}{6}
                \Biggr) \GA(10;t)
                -\frac{1}{6} \GA(4,2;t)
                +\frac{1}{24} \GA(6,2;t)
 +\frac{7}{8} \GA(8,2;t)
\nonumber\\ &&                
                +\frac{1}{4} \GA(8,1,2;t)
        \Biggr]
[-\gx_2(t)
                +(8+t) \gx_2'(t)
        ] 
+\frac{(1-t)}{2 t} \HA_{0,0,1}(t)
+\frac{\big(
        16-49 t+9 t^2\big) \zeta_2}{12 t}
\nonumber\\ && 
+\zeta_3
 + O(\ep),
\\
\F_3(t) &=& \frac{1}{\ep^2} \left[\frac{10}{3} - \frac{t}{6}\right] 
+ \frac{1}{\ep} \Biggl[-\frac{31}{6} + \frac{3 t}{8} - \left(\frac{1}{3} - \frac{1}{6 t} 
- \frac{t}{6}\right) \HA_1(t)
\Biggr] + 
\Biggl[ 
        \frac{3}{4} \ln(2) \gx_1(t)
\nonumber\\ &&
        +\frac{1}{12} \big(
                10+\pi  (-3 i +\sqrt{3})
\big) \gx_1(t)
        -\frac{\gx_2(t)}{3}
        +\frac{25}{54} [\gx_1(t) \GA(13;t) - \gx_2(t) \GA(7;t)]
\nonumber\\ && 
        +\frac{28}{27} [\gx_2(t) \GA(8;t) - \gx_1(t) \GA(14;t)]
        +\frac{1}{3} [\gx_1(t) \GA(16;t) - \gx_2(t) \GA(10;t)] 
\Biggr] \zeta_2
+\frac{31}{8} \ln(2) \gx_1(t)
\nonumber\\ && 
+\frac{1}{72} \big(
        265+31 \pi  (-3 i + \sqrt{3})
        \big) \gx_1(t)
-\frac{7}{2} \zeta_3 \gx_1(t)
-\frac{31 \gx_2(t)}{18}
+\frac{31}{18} [\gx_1(t) \GA(16;t) 
\nonumber\\ &&
- \gx_2(t) \GA(10;t)]
+\frac{7}{12} [\gx_1(t) \GA(5;t) - \gx_2(t) \GA(4;t)]
+\frac{655}{324} [\gx_1(t) \GA(13;t) - \gx_2(t) \GA(7;t)]
\nonumber\\ &&
+\frac{518}{81} [\gx_2(t) \GA(8;t) -  \gx_1(t) \GA(14;t)]
+\frac{1}{3} [\gx_1(t) \GA(5,2;t) - \gx_2(t) \GA(4,2;t)]
\nonumber\\ &&
+\frac{1}{12} [\gx_2(t) \GA(6,2;t) - \gx_1(t) \GA(12,2;t)]
+\frac{7}{4} [\gx_2(t) \GA(8,2;t) - \gx_1(t) \GA(14,2;t)]
\nonumber\\ && 
+\frac{1}{2} [\gx_2(t) \GA(8,1,2;t) - \gx_1(t) \GA(14,1,2;t)]
+ O(\ep).
\end{eqnarray}
The pole terms of the solutions are free of $_2F_1$--dependent letters both in $t$ and in $x$--space.
We checked numerically that the imaginary parts of $\F_1(t)$, $\F_2(t)$ and $\F_3(t)$ vanish for $t \in [0,1]$. 

We now transform to $x$--space via (\ref{eq:opt}) and obtain integral representations
in the physical region $x \in [0,1]$. 
The corresponding alphabet is obtained as a transformation of $\mathfrak{A}_2$. In these master integrals
only the cut in $t \in [1,\infty)$ contributes. Furthermore, regularizations 
at $x = 0,1$ are necessary in some cases. We first end up with a representation in terms of $\GA$--functions of $x$
and a number of special constants. At the point $x=1$ the $x$-- and $t$--expressions agree. 
Since the  expressions are rather voluminous, we will not show these 
expressions here but derive analytic expansions around $x = 0, 1/2$ and $x=1$, which have a more uniform 
structure. The corresponding series can be extended to very high orders.

Both the functions $\gx_{1,(2)}(t)$ are complex for $t > 1$. We replace $t =   1 + y$ and take the imaginary 
part. The transformation (\ref{eq:TRF}) introduces new constants given by $\GA$--functions at main argument 
one.
They can be calculated as described in Section~\ref{sec:42}. By expanding around $y = 0$ one can obtain the 
series 
expansion of the master integrals in the variable $1-x=y/(1+y)$. 
In general one expects the structure\footnote{In the numerical 
representations we normally use 20 digits.} 
\begin{eqnarray}
\sum_{k=-1}^\infty \sum_{l=0}^{L} \hat{a}_{k,l} (1-x)^k \ln^l(1-x).
\end{eqnarray}
In the present examples the logarithmic contributions do not contribute, cf.~(\ref{eq:F1c}--\ref{eq:F3c}).
One retains a number of terms by which a given precision in the region $x \in \left[1/2, 1\right]$ is 
obtained. 

In a similar way one proceeds to obtain an expansion around $x = 0$ and $x = 1/2$, respectively. For the 
associated differential 
equations the boundary conditions now known at $x=1$ are used to obtain the solutions 
around $x=0$ and $x = 1/2$. In both cases
new constants are contributing. They are at most two--fold integrals, cf.~Sect.~\ref{sec:42}, and are 
calculated numerically to high precision, in the 
cases they are no known numbers.

The series expansion around $x=0$ is given by
\begin{eqnarray}
\frac{1}{x} \sum_{k=0}^\infty \sum_{l=0}^{S} \hat{b}_{k,l} x^k \ln^l(x).
\end{eqnarray}
Here also $\GA$-constants at $x=1$ contribute.
Furthermore, we will need expansions around $x = 1/2$, 
\begin{eqnarray}
\sum_{k=0}^\infty  \hat{c}_{k,l} \left(\frac{1}{2} - x\right)^k
\end{eqnarray}
and further $\GA$-constants at $x=1/2$ contribute.
The expansion coefficients are given in Appendix~\ref{sec:C}.

One obtains
\begin{eqnarray}
\F_1(x) &=& \frac{8 x}{\ep^3} - \frac{1}{\ep^2} (2+9 x -  4 x \HA_0)
+ \frac{1}{\ep} \left[\frac{1}{12 x}[2+32 x+(71 +36 \zeta_2) x^2]
- \frac{1}{2} (2+9 x) \HA_0 + x \HA^2_0 \right]
\nonumber\\ && + \F_1^{(0)}(x) + O(\ep),
\\
\F_2(x) &=& - \frac{1}{\ep^2} 2 (1 - x) + \frac{1}{\ep} (1-x) \left[\frac{(1+8x)}{3x} - 
\HA_0(x)\right] + \F_2^{(0)}(x) + O(\ep),
\\
\F_3(x) &=& \frac{1}{\ep} \frac{(1 - x)^2}{6  x} + \F_3^{(0)}(x) + O(\ep).
\end{eqnarray}
For the expansion around $x = 1$ one obtains
\begin{eqnarray}
\label{eq:F1c}
\F_1^{(0), 1}(x) &=& \sum_{k=0}^\infty c_{1,k}^1 (1-x)^k.
\\
\F_2^{(0), 1}(x) &=& \sum_{k=1}^\infty c_{2,k}^1 (1-x)^k.
\\
\label{eq:F3c}
\F_3^{(0), 1}(x) &=& 
\sum_{k=2}^\infty c_{3,k}^1 (1-x)^k.
\end{eqnarray}
Correspondingly one obtains for the expansions around $x = 0$ and $x = 1/2$
\begin{eqnarray}
\label{eq:F1a}
\F_1^{(0), 0}(x) &=& 
c_{1,-1,1}^0 \frac{\ln{x}}{x} + c_{1,-1,0}^0 \frac{1}{x}
+ \sum_{k=0}^\infty [c_{1,k,0}^0 +  c_{1,k,1}^0 \ln(x) 
+ c_{1,k,2}^0 \ln^2(x) + c_{1,k,3}^0 \ln^3(x)] x^k,
\nonumber\\
\\
\F_2^{(0), 0}(x) &=&
c_{2,-1,1}^0 \frac{\ln{x}}{x} + c_{2,-1,0}^0 \frac{1}{x}
+ \sum_{k=0}^\infty [c_{2,k,0}^0 +  c_{2,k,1}^0 \ln(x) + c_{2,k,2}^0 \ln^2(x)] x^k,
\\
\label{eq:F3a}
\F_3^{(0), 0}(x) &=& 
c_{3,-1,1}^0 \frac{\ln{x}}{x} + c_{3,-1,0}^0 \frac{1}{x}
+ \sum_{k=0}^\infty [c_{3,k,0}^0 +  c_{3,k,1}^0 \ln(x) + c_{3,k,2}^0 \ln^2(x)] x^k,
\end{eqnarray}
and
\begin{eqnarray}
\label{eq:F1b}
\F_1^{(0), 1/2}(x) &=& \sum_{k=0}^\infty c_{1,k}^{1/2} \left(\frac{1}{2}-x\right)^k,
\\
\F_2^{(0), 1/2}(x) &=& \sum_{k=0}^\infty c_{2,k}^{1/2} \left(\frac{1}{2}-x\right)^k,
\\
\label{eq:F3b}
\F_3^{(0), 1/2}(x) &=& 
\sum_{k=0}^\infty c_{3,k}^{1/2} \left(\frac{1}{2}-x\right)^k.
\end{eqnarray}
After the transformation (\ref{eq:TRF}) is performed, the expressions for the master integrals contain
a series of constants. They can be calculated as $\GA$--functions numerically. The Mellin moments of the 
master integrals are given as $\zeta$--values, which have been calculated by different methods 
\cite{Blumlein:2017wxd} up to $N = 2000$. These provide further numerical precision tests.
We computed from the obtained $x$--space representations the first 10 Mellin moments, of the master 
integrals, and agree. 
Furthermore, we have compared the analytic results to numerical results in $x$--space which we
obtained by solving the associated first--order system of differential equations numerically with
the method applied in Ref.~\cite{Fael:2022miw} and found agreement.
\subsection{Iterating on the \boldmath $_2F_1$--solutions at first order}
\label{sec:42}

\vspace*{1mm} 
\noindent
After having solved all non--first--order--factorizing cases in analytic form, the other master integrals 
contributing 
to the system spanning a physical problem are of first order and can  now be integrated, since the respective 
inhomogeneities are successively obtained. At every order one has to solve an equation of the following form 
\begin{eqnarray}
y^{(1)}(t) + r(t) y(t) = h(t),
\end{eqnarray}
yielding \cite{KAMKE}
\begin{eqnarray}
\label{eq:DE1}
y(t) = \exp\left(-\int dt r(t)\right) \Biggl[C + \int h(t) \exp\left(\int dt r(t)\right) dt \Biggr]. 
\end{eqnarray}
The constant $C$ is fixed inserting a special value for $t$.
Since the rational functions can be partial fractioned 
allowing for complex constants  the 
exponential factors in (\ref{eq:DE1})
will become rational functions again. In the case of 
Kummer--Poincar\'e iterated letters \cite{KUMMER,POINCARE,LADAN,CHEN,GONCHAROV}
for $r(x)$ one obtains
\begin{eqnarray}
\label{eq:DE2}
y(t) = \frac{1}{t-a} \left[C  +
\int dt h(t) (t-a)
\right].
\end{eqnarray}
In the massive OME $A_{Qg}^{(3)}$ the master integrals outside of the
   two sectors that are related to ${}_2F_1$ solutions all fulfill
   first-order-factorizing differential equations of the form
\begin{eqnarray}
y'(x) + \frac{A}{x-b} y(x) = h(x),
\end{eqnarray}
which have the solution
\begin{eqnarray}
y(x) = (b-x)^{-A} \Biggl[
C b^A + \int_0^x dy (a-y)^A h(y)
\Biggr].
\end{eqnarray}
For half--integer constants $A$ one obtains root--valued letters, correspondingly.
The inohomogeneity $h(t)$ has itself an (iterated) integral representation down to 
the $_2F_1$--solutions. The further iteration adds one more iterated letter to 
the $\GA$--function from  the left.

As we saw above, in the present case the $_2F_1$--type letters appear in the $\GA$ 
index words next to each other, while, otherwise, letters are contributing forming 
iterated integrals. E.g. in the case of Kummer--Poincar\'e letters one may write
their iterated integral from the right. Accordingly, one may partially integrate from 
the left. The result is then a linear combination of two--fold integrals. 
As an example, let us consider the integral
\begin{eqnarray}
\Phi(x) &=& \GA(\{2,\Phi_1,\Phi_2,1,2\};x)
\nonumber\\ &=&
\int_0^x \frac{dx_1}{1-x_1}\int_0^{x_1} dx_2 \Phi_1(x_2) \int_0^{x_2} dx_3 \Phi_2(x_3) 
\int_0^{x_3} \frac{dx_4}{x_4}
\int_0^{x_4} \frac{dx_5}{1-x_5} 
\nonumber\\ &=&
\int_0^x \frac{dx_1}{1-x_1}\int_0^{x_1} dx_2 \Phi_1(x_2) \int_0^{x_2} dx_3 
\Phi_2(x_3) \Li_2(x_3)
\nonumber\\ 
&=& - \ln(1-x) \int_0^x dx_1 \Phi(x_1) \int_0^{x_1} dx_2 \Phi(x_2) \Li_2(x_2)
\nonumber\\ & & 
+ \int_0^x dx_1 \ln(1-x_1) \Phi(x_1) \int_0^{x_1} dx_2 \Phi(x_2) \Li_2(x_2).
\end{eqnarray}
Here the functions $\Phi_{1(2)}(x)$ denote the respective $_2F_1$--letters.
The transformation $t \rightarrow 1/x$ and the series expansion around $x=1$ will 
introduce a series of constants $\GA(a_1,...,a_k;1)$.
To compute them, the previously discussed integral representations can be used for the 
numerical integration, provided the numerical representations of the 
respective iterated integrals are known, cf.~e.g.~\cite{HPL,Vollinga:2004sn,Ablinger:2018sat}.
This representation holds up to the terms $O(\ep^0)$. More involved structures will appear in 
higher--order terms in $\ep$ for the master integrals.
\section{Numerical Representations} 
\label{sec:5}

\vspace*{1mm} 
\noindent
In the following we would like to make some brief remarks on possible numerical representations of the different
functions we discussed. For harmonic polylogarithms there are efficient numerical programs to high weight,
cf.~\cite{HPL,Vollinga:2004sn,Ablinger:2018sat}. Generalized harmonic polylogarithms can be calculated using the 
H\"older convolution  \cite{Borwein:1999js}, cf.~\cite{Vollinga:2004sn}. In some applications, 
cf.~\cite{Ablinger:2019etw,Ablinger:2014nga}, the generalized harmonic 
polylogarithms can be grouped to HPLs $\HA_{\vec{a}}(1-2x)$ in the final result.\footnote{In other 
applications, e.g. in massive QED, different but similar objects contribute, cf.~\cite{Ablinger:2020qvo}.} 
As we have seen in 
Section~\ref{sec:33}, for individual integrals Heaviside functions occur in $x$--space. They relate different 
$+$--functions to their Mellin transform. In Ref.~\cite{Ablinger:2014nga} the respective contributions canceled 
in the final (physical) expression, such that the Mellin transform is the usual one on support $x \in [0,1]$.

There are also codes for cyclotomic harmonic polylogarithms \cite{Ablinger:2018sat}. They can also be transformed 
into generalized harmonic polylogarithms using complex representations. In the case of the emergence of root--valued 
letters one will first try to rationalize as much as possible \cite{RATIONALIZE,Blumlein:2020jrf,RATRAAB}. This can 
also be done using 
procedures of {\tt HarmonicSums}. However, normally some of the root--valued structures will remain. Moreover,
the contributing iterated integrals may be numerous over longer alphabets, cf. e.g. \cite{Ablinger:2022wbb}. In this 
case one may first separate eventual distribution--valued terms. The remaining regular term, to be calculated 
for the interval $x \in [0,1]$, can be analytically expanded into Taylor series expansions modulated by logarithmic 
terms around $x = 0$ and $x=1$, to high precision. This also requires the power series expansion of the analytic 
continuation of the letters depending on $\tilde{g}_{1,(2)}(t)$ around $x = 0$ and 1.
In general, depending on the convergence radius of these series, further series
expansions inside the interval $[0,1]$ may become necessary.

\section{Conclusions} 
\label{sec:6}

\vspace*{1mm} 
\noindent
We have devised an algorithm to compute the inverse Mellin transform to Bjorken $x$--space directly 
from the resummation of the local operators from even or odd values of Mellin $N$, respectively, into propagators 
containing a continuous auxiliary variable $t \in \mathbb{R}$. The differential equations for the 
master integrals in this variable are either solved in terms of iterated or iterated non--iterative 
integrals. The results in Bjorken $x$--space are obtained by taking the imaginary part of the function 
after its analytic continuation $t \rightarrow \pm 1/x$. The latter operation can be performed by solving 
the differential equations  for the  iterated non--iterative integrals. In the case of only iterated 
integrals, general analytic implementations exist for different classes of functions. 
At higher order in $\ep$ additional $_2F_1$ letters will appear in the $\GA$--functions.
The constants contributing in the final $x$--space expressions are $\GA$--functions at $x=1$, by expanding 
around 
$x=0$ and $\GA$--functions at $x = \xi$ by expanding around $x = \xi$,~$\xi \in [0,1]$. The expressions 
in Mellin space for fixed values of $N$ are obtained by formal Taylor expansions of the analytic results 
in the parameter $t$. We also discussed numerical representations in $x$--space. Our calculations
were checked against a series of Mellin moments of the master integrals, which were computed using 
different methods. The present method allows to calculate the small--$x$ behaviour of the  considered
quantities directly, which is not easily possible from the $N$--space expressions. On the other hand,
$N$--space expressions allow to extract the large--$x$ behaviour, provided the corresponding difference
equations can be solved analytically in the limit $N \rightarrow \infty$.

\appendix
\section{Details of the analytic continuation}
\label{sec:A}

\vspace*{1mm} 
\noindent
In the following we derive (\ref{eq:DISC1}, \ref{eq:DISC2}) by using the residue theorem 
and discuss the separation of the distribution--valued contributions in $x$--space.

By using the representation of the Mellin transform (\ref{eq:MEL}) one obtains the 
following relation between $\tilde{f}(t)$ and $f(x)$,
\begin{align}
  \tilde{f}(t)
    &= \int\limits_0^1 \rmd x' \, \frac{t}{1-t x'} f(x')
  \,.
\end{align}
Here we consider for $f(x)$ a regular function.
Upon inserting the relation $t=1/x$, we get
\begin{align}
  \tilde{f}\left(\frac{1}{x}\right)
    &= \int\limits_0^1 \rmd x' \, \frac{f(x')}{x - x'}
  \,.
\end{align}
\begin{figure}[H]
  \tikzset{
    on each segment/.style={
      decorate,
      decoration={
        show path construction,
        moveto code={},
        lineto code={
          \path [#1]
          (\tikzinputsegmentfirst) -- (\tikzinputsegmentlast);
        },
        curveto code={
          \path [#1] (\tikzinputsegmentfirst)
          .. controls
          (\tikzinputsegmentsupporta) and (\tikzinputsegmentsupportb)
          ..
          (\tikzinputsegmentlast);
        },
        closepath code={
          \path [#1]
          (\tikzinputsegmentfirst) -- (\tikzinputsegmentlast);
        },
      },
    },
    mid arrow/.style={postaction={decorate,decoration={
          markings,
          mark=at position .5 with {\arrow[#1]{stealth}}
        }}},
  }
  \begin{subfigure}{0.32\textwidth}
    \centering
    \begin{tikzpicture}[thick]
      \draw[->] (-0.3,0) -- (4.3,0);
      \draw[->] (0,-1) -- (0,1);
      \node (xprime) at (4,0.8) {$x'$};
      \draw (xprime.south east) -- (xprime.south west) -- (xprime.north west);
      \draw[very thick] (0,0.1) -- (0,-0.1) node[black,below left] {$0$};
      \draw[very thick] (4,0.1) -- (4,-0.1) node[black,below] {$1$};
      \fill (2.5,0.1) circle (1.5pt) node[above] {$x+\I\delta$};
      \fill (2.5,-0.1) circle (1.5pt) node[below] {$x-\I\delta$};
      \draw[blue,very thick,mid arrow] (0,0) -- (4,0);
    \end{tikzpicture}
    \caption{}
    \label{fig:pole-positions}
  \end{subfigure}
  \begin{subfigure}{0.32\textwidth}
    \centering
    \begin{tikzpicture}[thick]
      \draw[->] (-0.3,0) -- (4.3,0);
      \draw[->] (0,-1) -- (0,1);
      \node (xprime) at (4,0.8) {$x'$};
      \draw (xprime.south east) -- (xprime.south west) -- (xprime.north west);
      \draw[very thick] (0,0.1) -- (0,-0.1) node[black,below left] {$0$};
      \draw[very thick] (4,0.1) -- (4,-0.1) node[black,below] {$1$};
      \fill (2.5,0) circle (1.5pt) node[below left=2mm] {$x$};
      \draw[blue,very thick,postaction={on each segment={mid arrow}}]
        (0,-0.05) -- (2.25,-0.05) arc (180:360:0.25cm) -- (4,-0.05);
      \draw[red,very thick,postaction={on each segment={mid arrow}}]
        (4,0.05) -- (2.75,0.05) arc (0:180:0.25cm) -- (0,0.05);
    \end{tikzpicture}
    \caption{}
    \label{fig:deformed-contours}
  \end{subfigure}
  \begin{subfigure}{0.32\textwidth}
    \centering
    \begin{tikzpicture}[thick]
      \draw[->] (-0.3,0) -- (4.3,0);
      \draw[->] (0,-1) -- (0,1);
      \node (xprime) at (4,0.8) {$x'$};
      \draw (xprime.south east) -- (xprime.south west) -- (xprime.north west);
      \draw[very thick] (0,0.1) -- (0,-0.1) node[black,below left] {$0$};
      \draw[very thick] (4,0.1) -- (4,-0.1) node[black,below] {$1$};
      \fill (2.5,0) circle (1.5pt) node[below left=2mm] {$x$};
      \draw[blue,very thick,postaction={on each segment={mid arrow}}] (2.75,0) arc (0:360:0.25cm);
    \end{tikzpicture}
    \caption{}
    \label{fig:circle-contour}
  \end{subfigure}
  \caption{\sf Illustration of the integration contours involved in extracting
    $f(x)$ from $\tilde{f}(t)$: (a)~integration contour for $\tilde{f}(1/x)$ (blue)
    and the position of the poles in the discontinuity; (b)~equivalent deformed
    contours to compute the discontinuity of $\tilde{f}(1/x)$ (in blue for the
    first term and in red for the second term); (c)~effective integration
    contour for the discontinuity of $\tilde{f}(1/x)$.}
  \label{fig:contours}
\end{figure}
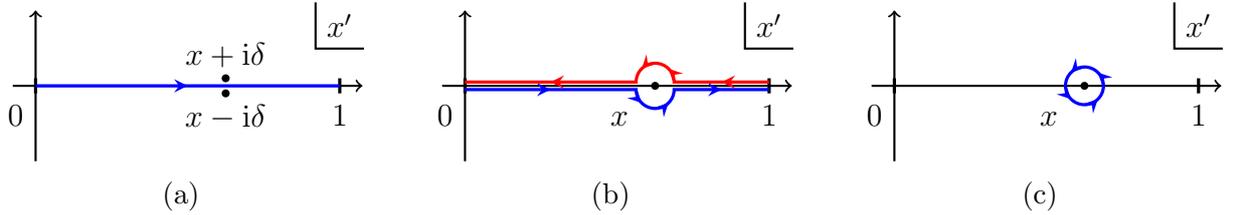
In order to extract $f(x)$ from $\tilde{f}(t=1/x)$, we can localize the integration
around the pole at $x' = x$ by calculating the discontinuity of $\tilde{f}$
across the branch cut induced by this pole,
\begin{align}
  \Disc_{x} \tilde{f}\left(\frac{1}{x}\right)
    &= \lim_{\delta \to 0^+} \left[
         \tilde{f}\left(\frac{1}{x+\I\delta}\right)
         -\tilde{f}\left(\frac{1}{x-\I\delta}\right)
       \right]
  \notag \\
    &= \lim_{\delta \to 0^+} \left[
         \int\limits_0^1 \rmd x' \, \frac{f(x')}{x+\I\delta - x'}
         -\int\limits_0^1 \rmd x' \, \frac{f(x')}{x-\I\delta - x'}
       \right].
\end{align}
The position of the poles in the first and second term is shown in
\cref{fig:pole-positions}. Equivalently, we can deform the integration contours
in the first and second term. The contour for the first term is shown in blue
and for the second term in red in \cref{fig:deformed-contours}. Since the
straight sections of the contours cancel out, only the circular contour shown in
\cref{fig:circle-contour} remains to be evaluated. Thus, we find with the help of
the residue theorem
\begin{align}
  \Disc_{x} \tilde{f}\left(\frac{1}{x}\right)
    &= \lim_{\delta \to 0} \oint\limits_{\mathclap{|x'-x| = \delta}} \rmd x' \,
       \frac{f(x')}{x - x'}
    = -2 \pi \I \, f(x)
  \,.
\end{align}
Note that the sign arises due to the form of the denominator.
Therefore, we can obtain $f(x)$ from $\tilde{f}(t)$ via
\begin{align}
  f(x) &= \frac{-1}{2 \pi \I}
    \Disc_x \tilde{f}\left(\frac{1}{x}\right)
  \,,
\end{align}
which leads to the relations (\ref{eq:DISC1}, \ref{eq:DISC2}).

We turn now to the separation of the distribution--valued contributions. We first 
consider the Mellin--transform of a typical distribution in $x$--space, $f(x),~x \in 
[0,1]$, occurring in QCD calculations, 
\begin{eqnarray}
\Mvec[f(x)](N) &=& 
\int_0^1 dx x^{N-1} f(x) 
\nonumber\\
&=& \int_0^1 dx x^{N-1} \Biggl[f_\delta \delta(1-x) + [f_+(x)]_+ 
+ f_{\rm reg,1}(x)
+ (-1)^{N-1} f_{\rm reg,2}(x)
\Biggr].
\end{eqnarray}
Here $f_+(x)$ is a linear combination of the functions $\HA_1^k(x)/(1-x),~~k \in \mathbb{N}$. The 
generating function in $t$--space is then obtained by
\begin{eqnarray}
\tilde{F}(t) &=& 
\int_0^1 dx' \Biggl\{
\frac{t}{1-t} \delta(1-x') f_\delta  
+ \left[\frac{t}{1-tx'} - \frac{t}{1-t}\right] f_+(x')
\nonumber\\ &&
+ \frac{t}{1-tx'} f_{\rm reg, 1}(x') 
+ \frac{t}{1+tx'} f_{\rm reg, 2}(x') 
\Biggr\}.
\end{eqnarray}
The distribution--valued parts can be integrated directly, cf.~(\ref{eq:dist1}--\ref{eq:dist5}), with
the first contributing $x$--space distributions and their $t$--space representation are given 
in Section~\ref{sec:21}. These contributions are subtracted from $\tilde{F}(t)$. One then obtains 
\begin{eqnarray}
\tilde{F}_{\rm reg}(t) = 
\int_0^1 dx' \Biggl[
\frac{t}{1-tx'} f_{\rm reg, 1}(x')
+ \frac{t}{1+tx'} f_{\rm reg, 2}(x')\Biggr].
\end{eqnarray}
$f_{\rm reg, 1}(x)$ and $f_{\rm reg, 2}(x)$ are reconstructed by forming
\begin{eqnarray}
\frac{1}{\pi} {\sf Im} \tilde{F}_{\rm reg, 1}\left(t = \frac{1}{x - i0} \right) &=& 
\frac{1}{\pi} {\sf Im}\int_0^1 dx' 
~\frac{1}{x - x' - i0}~f_{\rm reg, 1}(x') 
= f_{\rm reg, 1}(x),
\\
- \frac{1}{\pi} {\sf Im} \tilde{F}_{\rm reg, 2}\left(t = - \frac{1}{x - i0} \right) &=& 
 \frac{1}{\pi} {\sf Im}\int_0^1 dx' 
~ \frac{1}{x - x' - i0}~f_{\rm reg, 2}(x') 
= f_{\rm reg, 2}(x),
\end{eqnarray}
with $x \in [0,1]$.

\section{The solution after first decoupling for ${\sf F}_1(t)$} 
\label{sec:B}

\vspace*{1mm} 
\noindent
If one decouples the system of differential equations (\ref{eq:DIEQS}) for ${\sf F}_1(t)$ 
the solution of Eq.~(\ref{eq:HEUN1}) up to $O(1/\ep)$ is obtained as follows. 
For the homogeneous differential equation in the limit $\ep \rightarrow 0$ one obtains after the substitution 
$\F_1(t) = f_1(t)/t$
\begin{eqnarray}
\label{eq:HEUN1}
f_1^{(3)}(t) - \frac{2(4+5 t)}{t(1-t)(8+t)} f^{(2)}_1(t) + \frac{4}{t(1-t)(8+t)} f^{(1)}_1(t) = 0
\end{eqnarray}
and
\begin{eqnarray}
\label{eq:HEUN2}
\F_2(t) &=& 
\frac{342-105 t-t^2}{12 t}
+\frac{(1-t) (9+2 t) \HA_1(t)}{2 t^2}
+\frac{2 (1-t) \HA_{0,1}(t)}{t^2}
+\frac{6 \zeta_2}{t}
-\frac{(1-t) \F_1(t)}{t}
\nonumber \\ &&
-(1-t) \F_1'(t)
~,
\\
\F_3(t) &=& 
-\frac{54+111 t+52 t^2+3 t^3}{24 t^2}
-\frac{(1-t)^2 (-5+2 t) \HA_1(t)}{4 t^3}
+\frac{(1-t)^2 \HA_{0,1}(t)}{t^3}
-\frac{3 \zeta_2}{2 t}
\nonumber \\ &&
+\frac{(1-t)^2 \F_1'(t)}{t}
+\frac{1}{2} (1-t)^2 \F_1''(t)
~,
\end{eqnarray}
if one decouples for $\F_1(t)$ first.

We consider the homogeneous solution of the second--order differential equation in $g(t) = f^{(1)}(t)$ in the 
limit
$\ep \rightarrow 0$. 
The initial conditions are provided by the moments of the corresponding master integral,
to which the Taylor expansions around $t=0$ have to match.

Eq.~(\ref{eq:HEUN1}) is a Heun differential equation \cite{HEUN}, which has the following 
$_2F_1$--solutions 
\begin{eqnarray}
\label{eq:HEUN3a}
g_1(t) &=& 
i 2 \sqrt{\sqrt{3} \pi}
\frac{t^2 (8+t)^2}{(4-t)^4} 
\pFq{2}{1}{\frac{4}{3},\frac{5}{3}}{2}{z(t)},
\\
\label{eq:HEUN3b}
g_2(t) &=& i 2 \sqrt{\sqrt{3} \pi}
\frac{t^2 (8+t)^2}{(4-t)^4} \pFq{2}{1}{\frac{4}{3},\frac{5}{3}}{2}{1-z(t)},
\end{eqnarray}
with
\begin{eqnarray}
z(t) = \frac{27 t^2}{(4 - t)^3},
\end{eqnarray}
cf.~Ref.~\cite{Ablinger:2017bjx}.\footnote{The structure of (\ref{eq:HEUN3a}, \ref{eq:HEUN3b}) follows 
due to the relation $\alpha + \beta + 1 = 2\gamma; \alpha, \beta > 0$ for the corresponding $_2F_1$ function 
(\ref{eq:2F1}). We thank C.G.~Raab for this remark.}
For the analytic continuations to be carried out in the following it is very important to have closed
form solutions, such as the above $_2F_1$--solutions at hand.

The Wronski determinant \cite{WRONSKI} to a differential equation
\begin{eqnarray}
\label{eq:diffN}
y^{(n)}(t) 
+ p_1(t) y^{(n-1)}(t) 
+ p_2(t) y^{(n-2)}(t) \ldots 
+ p_n(t) y(t)  =  0
\end{eqnarray}
is given by
\begin{eqnarray}
W(t) = W(0) \exp\left[-\int_0^t p_1(t)\right] = \left|\begin{array}{ccc} 
y_1(t) & \ldots & y_n(t) \\
\vdots &        & \vdots \\
y_1^{(n-1)}(t) & \ldots & y_n^{(n-1)}(t) \end{array} \right|,
\end{eqnarray}
where $y_i(t)$ are the $n$ independent solutions of (\ref{eq:diffN}). The Wronskian of the solutions 
(\ref{eq:HEUN3a}, \ref{eq:HEUN3b}) reads
\begin{eqnarray}
W(t) = \frac{t(8+t)}{(1-t)^2}.
\end{eqnarray}
One may reduce higher--order derivatives of $g_{1,(2)}(t)$ by using their
differential equations. One thus obtains  
combinations of $g_{1,(2)}$ and $g_{1,(2)}'(t)$. Furthermore, one 
has
\begin{eqnarray}
g_1'(t) &=&  i 3^{1/4} \sqrt{\pi} \Biggl[64 \frac{t (2 + t) (8 + t)}{(4-t)^5} 
\pFq{2}{1}{\frac{4}{3},\frac{5}{3}}{2}{z(t)}
+  60 \frac{t^3 (8 + t)^3}{(4-t)^8} \pFq{2}{1}{\frac{7}{3},\frac{8}{3}}{3}{z(t)}\Biggr],
\\
g_2'(t) &=& i 3^{1/4} \sqrt{\pi} \Biggl[64\frac{t (2 + t) (8 + t)}{(4-t)^5}
\pFq{2}{1}{\frac{4}{3},\frac{5}{3}}{2}{1-z(t)}
+ 60 \frac{t^3 (8 + t)^3}{(4-t)^8} \pFq{2}{1}{\frac{7}{3},\frac{8}{3}}{3}{1-z(t)}
\Biggr].
\nonumber\\ 
\end{eqnarray}
The above solutions have already been calculated in 
Ref.~\cite{Ablinger:2017bjx}, up to a factor $i x^2/\sqrt{2}$, 
by changing variables to
\begin{eqnarray}
t \rightarrow 1 - \frac{9}{x^2}.
\end{eqnarray}
One may relate the latter functions further to complete elliptic integrals of the first and second kind,
${\bf K}(z_1(x)), {\bf K}(1-z_1(x)), {\bf E}(z_1(x))$ and ${\bf E}(1-z_1(x))$, with $z_1(x) = 
-16 z^3/[(x+1) (x-3)^3]$, as has been outlined 
in Ref.~\cite{Ablinger:2017bjx} in detail, by transforming the hypergeometric functions and using triangle 
relations~\cite{TAKEUCHI,IVH}. Here the particular structure of the function $z(t)$ has a deeper meaning
in the modular structure of these solutions, cf.~\cite{Ablinger:2017bjx}.
The solutions in terms of complete elliptic integrals have been applied in the first analytic calculation of 
the three--loop $\rho$--parameter of the Standard Model  
\cite{Blumlein:2018aeq}, which had been calculated semi--analytically in \cite{Grigo:2012ji} before.\footnote{Later in 
\cite{Abreu:2019fgk}, the results of \cite{Ablinger:2017bjx,Blumlein:2018aeq} have been confirmed.}
The emergence of the $_2F_1$--solutions in the present context is related to contributions of the so--called two--loop 
massive sun--rise integral, related also to the kite--integral, on which a very extensive literature exists. It dates 
back to \cite{SABRY}, cf. also~Refs.~\cite{ELL}.\footnote{For further references
see the extensive surveys given in Refs.~\cite{Blumlein:2019tmi,ELLREV}.}

In the present calculation we will use the $_2F_1$--representation (\ref{eq:HEUN3a}, \ref{eq:HEUN3b}) but not 
the representation due to complete elliptic integrals, since the number of higher transcendental functions is 
smaller and we would not really benefit from particular properties of the elliptic integrals.
We now transform the solutions (\ref{eq:HEUN3a}, \ref{eq:HEUN3b}) by $t \rightarrow 1/x$ for complex 
variables. One obtains
\begin{eqnarray}
G_1(x) &=& g_1\left(\frac{1}{t}\right) =
\frac{ i 2 \sqrt{\sqrt{3} \pi}(1 + 8 x)^2}{(1-4 x)^4} \pFq{2}{1}{\frac{4}{3},\frac{5}{3}}{2}
{- \frac{27 x}{(1 - 4 x)^3}},
\\
G_2(x) &=& g_2\left(\frac{1}{t}\right) 
= 
\frac{ i 2 \sqrt{\sqrt{3} \pi} (1 + 8 x)^2}{
(1-4 x)^4} \pFq{2}{1}{\frac{4}{3},\frac{5}{3}}{2}
{\frac{(1 - x) (1 + 8 x)^2}{(1 - 4 x)^3}}.
\end{eqnarray}
The integral representation of the hypergeometric function
\begin{eqnarray}
\pFq{2}{1}{\alpha,\beta}{\gamma}{z} = 
\frac{\Gamma(\gamma)}{\Gamma(\beta) \Gamma(\gamma-\beta)} \int_0^1 dt t^{\beta-1} 
(1-t)^{\gamma-\beta-1}(1-zt)^{-\alpha},~~{\sf Re}(\gamma) > {\sf Re}(\beta) > 0
\label{eq:2F1}
\end{eqnarray}
shows that $G_1(x)$ is purely imaginary for $x \in \left[0,\tfrac{1}{4}\right]$,
while this is the case for $G_2(x)$ for  $x \in \left[\tfrac{1}{4},1\right]$. 
At the boundaries one obtains 
\begin{eqnarray}
{\sf Re} \, G_2(0) &=& {\sf Re} \, G_1(1) =  - 3^{3/4} \sqrt{\pi},
\\
{\sf Im} \, G_1(0) &=& {\sf Im} \, G_2(1) =2 \sqrt{\sqrt{3} \pi},
\\
\lim_{x \rightarrow (1/4)^-} {\sf Re} \, G_1\left(x\right) &=& 
\lim_{x \rightarrow (1/4)^+} {\sf Re} \, G_2\left(x\right) = 0,
\\
\lim_{x \rightarrow (1/4)^+} {\sf Re} \, G_1\left(x\right) &=& 
\lim_{x \rightarrow (1/4)^-} {\sf Re} \, G_2\left(x\right) = -\frac{3^{3/4}}{2^{1/3} \pi^{3/2}} 
\Gamma^3\left(\frac{1}{3}\right),
\\
\lim_{x \rightarrow (1/4)^-} {\sf Im} \, G_1\left(x\right) &=& 
\lim_{x \rightarrow (1/4)^+} {\sf Im} \, G_2\left(x\right) = \frac{2^{2/3} 3^{1/4}}{\pi^{3/2}} \Gamma^3\left(
\frac{1}{3}\right),
\\
\lim_{x \rightarrow (1/4)^+} {\sf Im} \, G_1\left(x\right) &=& 
\lim_{x \rightarrow (1/4)^-} {\sf Im} \, G_2\left(x\right) = -\frac{3^{1/4}}{2^{1/3} \pi^{3/2}} 
\Gamma^3\left(\frac{1}{3}\right),
\end{eqnarray}
with the new constant $\Gamma(1/3)$.
The functions $G_{1(2)}(x)$ are discontinuous at $x = 1/4$ and have the following behaviour around $x=1$ and
$x=0$, respectively,
\begin{eqnarray}
{\sf Im} \, G_1(x) &=&  \frac{3^{3/4}}{\sqrt{\pi}} \left[- \frac{3}{2} \frac{1}{1-x}
+ \ln(1-x) \right] 
-\frac{3^{3/4}}{2 \pi} [-3 + 4 \ln(3)]
+ O((1-x)^1),
\\
{\sf Im} \, G_2(x) &=&  \frac{3^{3/4}}{\sqrt{\pi}}\left[-\frac{1}{6 x} + \ln(x)\right] + \frac{1}{
3^{1/4} \, 2 \sqrt{\pi}} + O(x^1).
\end{eqnarray}
The discontinuities disappear again in the inhomogeneous solutions, cf. also Ref.~\cite{Ablinger:2017bjx}.

Let us now go back to the $t$--space representation and solve the three inhomogeneous differential equations 
for ${\sf F}_k(t)$. 
The following alphabet contributes
\begin{eqnarray}
\label{eq:ALPH1}
\mathfrak{A}_1 &=& 
\left\{1,2,a_1,...,a_{16}\right\} =  \Biggl\{
\frac{1}{t}, 
\frac{1}{1-t},
g_1(t),
\frac{g_1(t)}{t},
\frac{g_1(t)}{1-t},
\frac{g_1(t)}{8+t},
\frac{g_1'(t)}{t},
\frac{g_1'(t)}{1-t},
\frac{g_1'(t)}{8+t},
\frac{g_1''(t)}{t},
g_2(t),
\nonumber \\ &&
\frac{g_2(t)}{t},
\frac{g_2(t)}{1-t},
\frac{g_2(t)}{8+t},
\frac{g_2'(t)}{t},
\frac{g_2'(t)}{1-t},
\frac{g_2'(t)}{8+t},
\frac{g_2''(t)}{t}
\Biggr\}.
\end{eqnarray}
We obtain for ${\sf F}_1(t)$ up to $O(\ep^{-1})$
\begin{eqnarray}
\lefteqn{{\sf F}_1(t) =} \nonumber\\ && 
\frac{8}{\ep^3} \left[1 + \frac{1}{t} \HA_1(t)\right]
+ \frac{1}{\ep^2} \Biggl[
- \frac{1}{6} (106 + t) - \frac{9 + 2 t}{t} \HA_1(t) - \frac{4}{t} \HA_{0,1}(t)
\Biggr]
\nonumber\\ &&
+ \frac{1}{\ep~t} \Biggl\{
\frac{1}{128(1-t)} \Biggl[
-2654 t
+\biggl(
        2302
        -44 t
        -224 (-1+t) \HA_{0,1}(t)
\biggr) \HA_1(t)
-95 (1-t) \HA_1(t)^2
\nonumber \\ &&
+16 (36+t) \HA_{0,1}(t)
+256 \HA_{0,0,1}(t)
-256 t \HA_{0,0,1}(t)
-448 \HA_{0,1,1}(t)
+448 t \HA_{0,1,1}(t)
\Biggr]
\nonumber \\ &&
+ i \Biggl[
-\frac{1}{96\sqrt[4]{3} \sqrt{\pi } } \big(
        1109
        +27 \ln(2) \big(
                125
                +24 \zeta_2
        \big)
        +144 \zeta_2
\big) \GA(a_1;t)
-\frac{1}{32} \big(
        125
        \nonumber \\ &&
        +24 \zeta_2
\big) \sqrt[4]{3} \sqrt{\pi } \GA(a_9;t)
\Biggr]
+
\frac{1}{64} \biggl(
        161
        +18 \zeta_2
\biggr) \GA(a_1,a_{10};t)
+\frac{11539 \GA(a_1,a_{11};t)}{20736}
\nonumber \\ &&
-\biggl(
        \frac{33713}{20736}
        +\frac{9 \zeta_2}{32}
\biggr) \GA(a_1,a_{12};t)
-\frac{269}{128} \GA(a_1,a_{13};t)
-\frac{733}{576} \GA(a_1,a_{14};t)
\nonumber \\ &&
-\biggl(
         \frac{23939}{1152}
        +\frac{9 \zeta_2}{4}
\biggr) \GA(a_1,a_{15};t)
-\frac{1}{64} \big(
        161
        +18 \zeta_2
\big) \GA(a_9,a_2;t)
-\frac{11539 \GA(a_9,a_3;t)}{20736}
\nonumber \\ &&
+\biggl(
        \frac{33713}{20736}
        +\frac{9 \zeta_2}{32}
\biggr) \GA(a_9,a_4;t)
+\frac{269}{128} \GA(a_9,a_5;t)
+\frac{733}{576} \GA(a_9,a_6;t)
+\biggl(
        \frac{23939}{1152}
        \nonumber \\ &&
        +\frac{9 \zeta_2}{4}
\biggr) \GA(a_9,a_7;t)
+\frac{12845 \GA(a_1,a_{10},2;t)}{18432}
+\frac{371}{648} \GA(a_1,a_{11},2;t)
-\frac{20629 \GA(a_1,a_{12},2;t)}{165888}
\nonumber \\ &&
-\frac{283}{128} \GA(a_1,a_{13},2;t)
-\frac{371}{144} \GA(a_1,a_{14},2;t)
-\frac{4315 \GA(a_1,a_{15},2;t)}{2304}
-\frac{43}{64} \GA(a_1,a_{16},2;t)
\nonumber \\ &&
-\frac{12845 \GA(a_9,a_2,2;t)}{18432}
-\frac{371}{648} \GA(a_9,a_3,2;t)
+\frac{20629 \GA(a_9,a_4,2;t)}{165888}
+\frac{283}{128} \GA(a_9,a_5,2;t)
\nonumber \\ &&
+\frac{371}{144} \GA(a_9,a_6,2;t)
+\frac{4315 \GA(a_9,a_7,2;t)}{2304}
+\frac{43}{64} \GA(a_9,a_8,2;t)
+\frac{137}{512} \GA(a_1,a_{10},1,2;t)
\nonumber \\ &&
+\frac{37}{162} \GA(a_1,a_{11},1,2;t)
-\frac{1625 \GA(a_1,a_{12},1,2;t)}{41472}
-\frac{133}{128} \GA(a_1,a_{13},1,2;t)
-\frac{37}{36} \GA(a_1,a_{14},1,2;t)
\nonumber \\ &&
-\frac{85 \GA(a_1,a_{15},1,2;t)}{1152}
-\frac{137}{512} \GA(a_9,a_2,1,2;t)
-\frac{37}{162} \GA(a_9,a_3,1,2;t)
+\frac{1625 \GA(a_9,a_4,1,2;t)}{41472}
\nonumber \\ &&
+\frac{133}{128} \GA(a_9,a_5,1,2;t)
+\frac{37}{36} \GA(a_9,a_6,1,2;t)
+\frac{85 \GA(a_9,a_7,1,2;t)}{1152}
\Biggr\} 
+ O(\ep^0).
\label{eq:LONG}
\end{eqnarray}
It is the first order in which the homogeneous $_2F_1$--solutions seems to contribute. 
Here we refer to the 
letters of alphabet $\mathfrak{A}_1$, Eq.~(\ref{eq:ALPH2}),
and up to depth four $\GA$--functions, containing $_2F_1$--letters contribute. The expression reduces, however, to
(\ref{eq:F1A}) for the pole terms, if one first decouples for ${\sf F}_3(t)$, which is difficult to see a 
posteriori. We have compared the first ten Taylor coefficients 
of both representations and they agree. In (\ref{eq:LONG}) even some HPLs emerge, which are not 
present in (\ref{eq:F1A}).
\section{The expansion coefficients of series representations} 
\label{sec:C}

\vspace*{1mm} 
\noindent
The first expansion coefficients in Eqs.~(\ref{eq:F1c}--\ref{eq:F3c})
are given by
\begin{align}
  c_{1,0}^1  &= -11.16958740964 \,, &
  c_{1,1}^1  &=  2.109346617266 \,,
  \nonumber \\
  c_{1,2}^1  &=  0.936851756584 \,, &
  c_{1,3}^1  &=  0.286064880707 \,,
  \nonumber \\
  c_{1,4}^1  &=  0.127032314586 \,, &
  c_{1,5}^1  &=  0.063499317190 \,,
  \nonumber \\
  c_{1,6}^1  &=  0.034750073376 \,, &
  c_{1,7}^1  &=  0.021455163556 \,,
  \nonumber \\
  c_{1,8}^1  &=  0.015822146627 \,, &
  c_{1,9}^1  &=  0.014262405540 \,,
  \nonumber\\
  c_{1,10}^1 &=  0.014967991102 \,,
\end{align}
\begin{align}
  c_{2,1}^1  &= -2.217839692102 \,, &
  c_{2,2}^1  &= -0.718697587104 \,,
  \nonumber \\
  c_{2,3}^1  &= -0.370323781129 \,, &
  c_{2,4}^1  &= -0.189000503072 \,,
  \nonumber \\
  c_{2,5}^1  &= -0.084433691142 \,, &
  c_{2,6}^1  &= -0.016330161839 \,,
  \nonumber \\
  c_{2,7}^1  &=  0.031991333568 \,, &
  c_{2,8}^1  &=  0.068481112319 \,,
  \nonumber \\
  c_{2,9}^1  &=  0.097368528228 \,, &
  c_{2,10}^1 &=  0.121096539717 \,,
\end{align}
\begin{align}
  c_{3,2}^1  &=  0.390651206448 \,, &
  c_{3,3}^1  &=  0.322358345756 \,,
  \nonumber \\
  c_{3,4}^1  &=  0.295156359854 \,, &
  c_{3,5}^1  &=  0.281300038991 \,,
  \nonumber \\
  c_{3,6}^1  &=  0.273875311020 \,, &
  c_{3,7}^1  &=  0.270132738635 \,,
  \nonumber \\
  c_{3,8}^1  &=  0.268709892411 \,, &
  c_{3,9}^1  &=  0.268837838844 \,,
  \nonumber \\
  c_{3,10}^1 &=  0.270043649148 \,.
\end{align}
The coefficients of Eqs.~(\ref{eq:F1a}--\ref{eq:F3a}) read
\begin{align}
  c_{1,-1,1}^0 &= -\frac{1}{6} \,, &
  c_{1,-1,0}^0 &= -\frac{3}{4} \,, &
  c_{1,0,0}^0  &=  \frac{11}{4} - \frac{3}{4} \zeta_2 \,,
  \nonumber \\
  c_{1,0,1}^0  &=  \frac{29}{6} \,, &
  c_{1,0,2}^0  &=  \frac{5}{4} \,, &
  c_{1,1,0}^0  &= -\frac{113}{16} - \frac{27}{8} \zeta_2 + 5 \zeta_3 \,,
  \nonumber \\
  c_{1,1,1}^0  &=  \frac{83}{24} + \frac{3}{2} \zeta_2 \,, &
  c_{1,1,2}^0  &= -\frac{3}{8} \,, &
  c_{1,1,3}^0  &= -\frac{5}{6} \,,
  \nonumber \\
  c_{1,2,0}^0  &= -\frac{79}{12} \,, &
  c_{1,2,1}^0  &=  3 \,, &
  c_{1,3,0}^0  &=  \frac{19}{4} \,,
  \nonumber \\
  c_{1,3,1}^0  &= -\frac{9}{4} \,, &
  c_{1,3,2}^0  &= -3 \,, &
  c_{1,4,0}^0  &= -\frac{7613}{720} \,,
  \nonumber \\
  c_{1,4,1}^0  &=  \frac{143}{12} \,, &
  c_{1,4,2}^0  &=  5 \,, &
  c_{1,5,0}^0  &=  \frac{64103}{2400} \,,
  \nonumber \\
  c_{1,5,1}^0  &= -\frac{891}{20} \,, &
  c_{1,5,2}^0  &= -18 \,,
\end{align}
\begin{align}
  c_{2,-1,1}^0 &= -\frac{1}{3} \,, &
  c_{2,-1,0}^0 &= -\frac{5}{4} \,, &
  c_{2,0,0}^0  &=  \frac{1}{2} - \frac{3}{4} \zeta_2 \,,
  \nonumber \\
  c_{2,0,1}^0  &=  \frac{13}{6} \,, &
  c_{2,0,2}^0  &=  \frac{5}{4} \,, &
  c_{2,1,0}^0  &=  \frac{1}{4} + \frac{3}{4} \zeta_2 \,,
  \nonumber \\
  c_{2,1,1}^0  &= -\frac{10}{3} \,, &
  c_{2,1,2}^0  &=  \frac{7}{4} \,, &
  c_{2,2,0}^0  &=  \frac{49}{12} \,,
  \nonumber \\
  c_{2,2,1}^0  &= -\frac{3}{2} \,, &
  c_{2,2,2}^0  &= -3 \,, &
  c_{2,3,0}^0  &= -\frac{65}{6} \,,
  \nonumber \\
  c_{2,3,1}^0  &=  \frac{27}{2} \,, &
  c_{2,3,2}^0  &=  6 \,, &
  c_{2,4,0}^0  &=  \frac{6493}{240} \,,
  \nonumber \\
  c_{2,4,1}^0  &= -\frac{225}{4} \,, &
  c_{2,4,2}^0  &= -21 \,, &
  c_{2,5,0}^0  &= -\frac{32837}{400} \,,
  \nonumber \\
  c_{2,5,1}^0  &=  \frac{5199}{20} \,, &
  c_{2,5,2}^0  &=  87 \,,
\end{align}
\begin{align}
  c_{3,-1,1}^0 &= -\frac{1}{6} \,, &
  c_{3,-1,0}^0 &= -\frac{3}{8} \,, &
  c_{3,0,0}^0  &=  \frac{1}{2} \,,
  \nonumber \\
  c_{3,0,1}^0  &= -\frac{7}{6} \,, &
  c_{3,1,0}^0  &=  \frac{9}{8} \,, &
  c_{3,1,1}^0  &=  \frac{7}{12} \,,
  \nonumber \\
  c_{3,1,2}^0  &= -\frac{3}{2} \,, &
  c_{3,2,0}^0  &= -\frac{13}{3} \,, &
  c_{3,2,1}^0  &=  6 \,, &
  \nonumber \\
  c_{3,2,2}^0  &=  3 \,, &
  c_{3,3,0}^0  &=  \frac{259}{24} \,, &
  c_{3,3,1}^0  &= -30 \,,
  \nonumber \\
  c_{3,3,2}^0  &= -\frac{21}{2} \,, &
  c_{3,4,0}^0  &= -\frac{451}{15} \,, &
  c_{3,4,1}^0  &=  153 \,,
  \nonumber \\
  c_{3,4,2}^0  &=  48 \,, &
  c_{3,5,0}^0  &=  \frac{7017}{80} \,, &
  c_{3,5,1}^0  &= -\frac{3369}{4} \,,
  \nonumber \\
  c_{3,5,2}^0  &= -249 \,.
\end{align}
The above rational constants have been determined using {\tt PSLQ} \cite{PSLQ}.
They do structurally agree with those of $a_{Qq}^{\rm PS, (3)}$ of
Ref.~\cite{Ablinger:2019etw}, which is related to $a_{Qg}^{\rm (3)}$ by color rescaling with 
$C_A/C_F$ in the 
leading term \cite{Catani:1990eg},
where $C_F = (N_C^2-1)/(2N_C), C_A = N_C$ and $N_C =3$ for Quantum 
Chromodynamics.\footnote{Similar analytic
patterns have been observed for the massive three--loop form factor \cite{FORMF}.}
In the expansion of ${\sf F}_3(x)$ no $\zeta$-terms seem to contribute for the first 100 terms in $x$, while
${\sf F}_2(x)$ depends on $\zeta_2$ and ${\sf F}_1(x)$ also on $\zeta_3$. The master integrals contributing to 
$a_{Qg}^{\rm (3)}$ may in principle also depend on $\zeta_4$ and ${\sf B}_4$, 
cf.~\cite{Bierenbaum:2009mv}, Eq.~(4.10). 

Finally, one obtains for the coefficients of Eqs.~(\ref{eq:F1b}--\ref{eq:F3b}) 
\begin{align}
  c_{1,0}^{1/2}  &= -9.834184787511 \,, &
  c_{1,1}^{1/2}  &= 3.355232766926  \,,
  \nonumber \\
  c_{1,2}^{1/2}  &= 1.701654239373  \,, &
  c_{1,3}^{1/2}  &= 0.933416116957  \,,
  \nonumber \\
  c_{1,4}^{1/2}  &= 0.891822658934  \,, &
  c_{1,5}^{1/2}  &= 1.440452967512  \,,
  \nonumber \\
  c_{1,6}^{1/2}  &= 3.207281678902  \,, &
  c_{1,7}^{1/2}  &= 7.783359303513  \,,
  \nonumber \\
  c_{1,8}^{1/2}  &= 18.79614079037  \,, &
  c_{1,9}^{1/2}  &= 44.28851410206  \,,
  \nonumber \\
  c_{1,10}^{1/2} &= 101.8245323374  \,,
\end{align}
\begin{align}
  c_{2,0}^{1/2}  &= -1.348611882678 \,, &
  c_{2,1}^{1/2}  &= -3.320927437135 \,,
  \nonumber \\
  c_{2,2}^{1/2}  &= -1.536412632474 \,, &
  c_{2,3}^{1/2}  &= -0.267319762707 \,,
  \nonumber \\
  c_{2,4}^{1/2}  &=  2.269831457716 \,, &
  c_{2,5}^{1/2}  &=  7.982990699375 \,,
  \nonumber \\
  c_{2,6}^{1/2}  &=  20.82740039869 \,, &
  c_{2,7}^{1/2}  &=  49.17055989829 \,,
  \nonumber \\
  c_{2,8}^{1/2}  &=  110.6955042191 \,, &
  c_{2,9}^{1/2}  &=  242.5826709616 \,,
  \nonumber \\
  c_{2,10}^{1/2} &=  522.6300919150 \,,
\end{align}
\begin{align}
  c_{3,0}^{1/2}  &= 0.173692073146 \,, &
  c_{3,1}^{1/2}  &= 0.986776221633 \,,
  \nonumber \\
  c_{3,2}^{1/2}  &= 2.415478375577 \,, &
  c_{3,3}^{1/2}  &= 4.469951985772 \,,
  \nonumber \\
  c_{3,4}^{1/2}  &= 8.772564418720 \,, &
  c_{3,5}^{1/2}  &= 17.62005543760 \,,
  \nonumber \\
  c_{3,6}^{1/2}  &= 35.78474174591 \,, &
  c_{3,7}^{1/2}  &= 73.07722039062 \,,
  \nonumber \\
  c_{3,8}^{1/2}  &= 149.6247109869 \,, &
  c_{3,9}^{1/2}  &= 306.6679998469 \,,
  \nonumber \\
  c_{3,10}^{1/2} &= 628.6136390924 \,.
\end{align}

\vspace*{5mm}
\noindent
{\bf Acknowledgment.} We would like to thank D.J.~Broadhurst, A.~De Freitas, P.~Marquard, C.G.~Raab, and 
C.~Schneider for discussions. This work has received funding in part from the European Research Council 
(ERC) under the European Union's Horizon 2020 research and innovation programme grant agreement 101019620 (ERC 
Advanced Grant TOPUP). 

{\footnotesize

\end{document}